\documentclass[aps,prl,twocolumn,preprintnumbers,10pt,nofootinbib]{revtex4-2}

\pdfoutput=1

\usepackage[letterpaper,top=1.45cm,bottom=1.65cm,left=2cm,right=2cm]{geometry}

\usepackage[T1]{fontenc} 
\usepackage{graphicx}
\usepackage{epsfig}
\usepackage{amssymb}
\usepackage{amssymb}
\usepackage{amsfonts}
\usepackage{amsmath,euscript,array,mathrsfs}
\usepackage{bbold}
\usepackage{epsf}
\usepackage{slashed}
\usepackage{comment}
\usepackage{colortbl}
\usepackage{enumitem}
\usepackage{float}

\usepackage[caption=false]{subfig}
\usepackage[colorlinks=true,linktocpage=true,linkcolor=blue,citecolor=blue]{hyperref}

\usepackage{tikz}
\usetikzlibrary{decorations.markings,decorations.pathmorphing}

\newcommand{\ydot}{\hspace{-0.5mm}\cdot \hspace{-0.5mm}}

\begin{document}
\onecolumngrid

\title{Relativistic Navier--Stokes description of the quark-gluon plasma radial flow}

\author{Yago Bea}

\affiliation{Departament de F\'\i sica Qu\`antica i Astrof\'\i sica and Institut de
	Ci\`encies del Cosmos (ICC), Universitat de Barcelona, Mart\'\i\ i Franqu\`es
	1, ES-08028, Barcelona, Spain}

\affiliation{Departamento de Matem\'atica Aplicada II, Universidade de Vigo,
	As Lagoas, 32004 Ourense, Spain}

\begin{abstract}

Using relativistic Navier--Stokes in the well-posed
Bemfica--Disconzi--Noronha--Kovtun (BDNK) formulation, 
with a QCD equation of state and transport coefficients, we perform simulations that describe
the radial flow in central Pb--Pb collisions measured at the LHC, 
providing a proof of principle for its viability in heavy-ion phenomenology.
We identify causal frames that retain the Landau-frame interpretation, and verify frame
robustness. M\"uller--Israel--Stewart simulations with identical input show
large second-order effects, exceeding the frame dependence of relativistic
Navier--Stokes, as well as causality violations.

\end{abstract}

\maketitle

\noindent
\textbf{1. Introduction.} 
Relativistic hydrodynamics has been remarkably successful in describing the quark--gluon plasma (QGP) created in heavy-ion collisions~\cite{Romatschke:2017ejr,Busza:2018rrf,Heinz:2004qz,Teaney:2009qa,Heinz:2024jwu},
where viscous effects prove relevant~\cite{Teaney:2003kp}.
The natural framework for incorporating these effects would be the
relativistic Navier--Stokes equations, but the original formulations by Eckart~\cite{Eckart:1940te} and Landau and Lifshitz~\cite{LandauBook} suffer from severe pathologies~\cite{Hiscock:1985zz}. This led to the adoption of the formulation of M\"uller, Israel and Stewart (MIS)~\cite{Muller:1967zza,Israel:1976tn,Israel:1979wp}, which alleviates these problems by introducing relaxation equations for the dissipative fluxes. 
MIS-based theories~\cite{Denicol:2012cn,Baier:2007ix} are now the standard framework in heavy-ion phenomenology and are implemented in widely used codes such as MUSIC~\cite{Schenke:2010nt,Schenke:2010rr,Paquet:2015lta}, SONIC~\cite{Habich:2014jna} and VISHNU~\cite{Shen:2014vra}.

Despite this success, MIS theories present limitations.
Their nonlinear causality and well-posedness conditions have been
derived~\cite{Bemfica:2020xym,Cordeiro:2026jle}, and checks in realistic
simulations of heavy-ion collisions have revealed significant
violations~\cite{Plumberg:2021bme,Chiu:2021muk,ExTrEMe:2023nhy,Domingues:2024pom}.
Since experimental data are well reproduced, the consequences may be mild,
but quantifying how these limitations affect the analysis of experimental
data remains an open problem.
This requires a formulation free of these limitations, such
as the one discussed next.

For decades, a consistent formulation of the relativistic Navier--Stokes equations remained elusive.
Significant progress was made by Bemfica, Disconzi and Noronha~\cite{Bemfica:2017wps,Bemfica:2019knx,Bemfica:2020zjp} and Kovtun~\cite{Kovtun:2019hdm} (BDNK), who derived the conditions that a hydrodynamic frame must satisfy for the equations to be well-posed and causal.
The Eckart and Landau frames do not satisfy these conditions, which explains the pathologies of the original formulations.
In the language of effective field theory (EFT), a change of frame corresponds to a field redefinition.
Following these mathematical results, several numerical
studies~\cite{Bea:2023rru,Bantilan:2022ech,Pandya:2021ief,Pandya:2022pif,Pandya:2022sff,Clarisse:2025lli,Keeble:2025bkc} provided evidence that the equations admit physically sensible solutions in simple and controlled setups.
More recently, they have been applied to neutron-star evolution~\cite{Shum:2025jnl}
and, while this work was under review, to boost-invariant 2+1-dimensional
systems with heavy-ion initial conditions, including a comparison with
MIS~\cite{Patel:2026dba}.

In this Letter, we perform boost-invariant, axially symmetric (1+1)-dimensional simulations of the relativistic Navier--Stokes equations with a QCD equation of state and transport coefficients, and use them to describe the radial flow (the collective outward expansion of the plasma) in central Pb--Pb collisions at the LHC.
To this end, we identify a class of frames that preserve the Landau-frame
interpretation of the hydrodynamic variables and in which the causality
conditions take a simple form, and construct explicit examples that are
causal for every state.
We also propose a prescription for the initial time derivatives, required
since the equations are second order in time.
Comparing evolutions in two different frames, we show that the results are
robust under the choice of frame.
Finally, we perform MIS simulations with the same input; an EFT analysis
shows that their second-order effects are much larger than the frame
dependence of relativistic Navier--Stokes, and they violate causality at early times.



\noindent
\textbf{2. Relativistic Navier--Stokes equations.} We consider a fluid with equation of state $p(\epsilon)$,
where $\epsilon$ is the energy density and $p$ the pressure, with speed of sound
$c_s^2=dp/d\epsilon$ and four-velocity $u^\mu$ satisfying $u^\mu u_\mu=-1$.
We define $\Delta^{\mu\nu}:=g^{\mu\nu}+u^\mu u^\nu$,
$\dot{\epsilon}:=u^\mu\nabla_\mu\epsilon$,
$\nabla_\perp^\mu:=\Delta^{\mu\nu}\nabla_\nu$,
$\nabla\ydot u:=\nabla_\mu u^\mu$ and
$\sigma^{\mu\nu}:=\nabla^\mu_\perp u^\nu+\nabla^\nu_\perp u^\mu
-\tfrac{2}{3}\Delta^{\mu\nu}\nabla\ydot u$.
To first order in the gradient expansion, the constitutive relations in the Landau frame are~\cite{LandauBook}
\begin{equation}
  T^{\mu\nu}=\epsilon\,u^\mu u^\nu+p\,\Delta^{\mu\nu}
  -\eta\,\sigma^{\mu\nu}-\zeta\,\nabla\ydot u\,\Delta^{\mu\nu}\,,
  \label{Constitutive_first_order_non_conformal_Landau}
\end{equation}
with $\eta$ and $\zeta$ the shear and bulk viscosities, and the dynamics follows from the conservation of the stress tensor,
\begin{equation}
  \nabla_\mu T^{\mu\nu}=0\,.
  \label{conservation0}
\end{equation}
For ideal hydrodynamics, $\eta=\zeta=0$, these equations reduce to
\begin{equation}
  \frac{\dot{\epsilon}}{\epsilon+p}+\nabla\ydot u=0\,,\qquad
  \dot{u}^\mu+c_s^2\,\frac{\nabla_\perp^\mu\epsilon}{\epsilon+p}=0\,.
  \label{EOMs_ideal_hydrodynamics}
\end{equation}

We adopt a perspective in which the hydrodynamic frames of interest are obtained from the Landau-frame constitutive relations~\eqref{Constitutive_first_order_non_conformal_Landau} through first-order field redefinitions, 
providing a direct connection with the frame convention commonly used in QGP phenomenology. 
In the spirit of EFT, we consider 
\begin{equation}
  \epsilon \rightarrow \epsilon+\mathcal{A}\,,\qquad
  u^\mu \rightarrow u^\mu+\frac{\mathcal{Q}^\mu}{\epsilon+p}\,,
  \label{Field_redefinition_non_conformal2}
\end{equation}
with
\begin{subequations}
\begin{align}
  \mathcal{A} &:= \big(\eta+\tfrac{3}{4}\zeta\big)
    \Big(a_1 \frac{\dot{\epsilon}}{\epsilon+p}+\tilde{a}_1 \nabla\ydot u\Big)\,,
    \label{Our_field_redefinitiona}\\
  \mathcal{Q}^{\mu} &:= a_2 \big(\eta+\tfrac{3}{4}\zeta\big)
    \Big(\dot{u}^{\mu}+c_s^2 \frac{\nabla_{\perp}^{\mu}\epsilon}{\epsilon+p}\Big)\,.
    \label{Our_field_redefinitionb}
\end{align}
\label{Our_field_redefinition}
\end{subequations}
Here $\{a_1,\tilde a_1,a_2\}$ are arbitrary functions of $\epsilon$ parametrizing the most general first-order field redefinition compatible with Poincar\'e symmetry and extensivity~\cite{Kovtun:2019hdm}, and the factor $(\eta+3\zeta/4)$ is introduced for convenience. 

Under the field redefinition~\eqref{Field_redefinition_non_conformal2}, the stress tensor~\eqref{Constitutive_first_order_non_conformal_Landau} becomes
\begin{align}
  T^{\mu\nu} &= \left(\epsilon + \mathcal{A}\right)u^\mu u^\nu
  + \left(p + c_s^2\mathcal{A}\right)\Delta^{\mu\nu} \nonumber\\
  &\quad + \mathcal{Q}^{\mu}u^\nu + u^{\mu}\mathcal{Q}^\nu
  - \eta\,\sigma^{\mu\nu} - \zeta\,\nabla\ydot u\,\Delta^{\mu\nu}\,,
  \label{stress_tensor_generic_frame}
\end{align}
where second- and higher-order terms in the gradient expansion are neglected. Substituting~\eqref{stress_tensor_generic_frame} into the
conservation equation~\eqref{conservation0} gives the \textit{relativistic Navier--Stokes equations} in a general frame~\cite{SM}. Each frame leads to a different system of partial differential equations, so that there is a family of relativistic Navier--Stokes equations, one for each field redefinition~\cite{Kovtun:2019hdm}.
The additional terms $\mathcal{A}$ and $\mathcal{Q}^\mu$ in the stress tensor~\eqref{stress_tensor_generic_frame} have three properties that are key for phenomenological applications:
\begin{itemize}[left=0pt]
  \item They do not change the physical content to first order.
  \item For the subset $\tilde a_1=a_1$, which we adopt in this work, they also preserve the Landau-frame interpretation of the hydrodynamic variables.
  \item They can be chosen to `regulate' the equations, so that they are well behaved.
\end{itemize}
We now explain these statements.

Since $\mathcal{A}$ and $\mathcal{Q}^\mu$ originate from a field
redefinition, they do not modify the physical content to first order in
the gradient expansion, as long as the system is in the EFT
regime~\cite{Kovtun:2019hdm}. For $\tilde a_1=a_1$, they are moreover
proportional to the ideal equations of
motion~\eqref{EOMs_ideal_hydrodynamics}, and hence of second order on
shell, i.e., upon using the viscous equations of motion. The redefined variables thus
differ from the Landau-frame ones only by terms that are negligible to
first order, so that $\epsilon$ and $u^\mu$ retain their Landau-frame
meaning.

This is not the case for general field redefinitions. With a chemical
potential, for instance, the Landau and Eckart frames assign different
meanings to the fluid velocity (the velocity of the energy and of the
charge, respectively), since the redefinition relating them is not
proportional to the ideal equations of motion. This motivates us to introduce the following classification:
\begin{itemize}[left=0pt]
  \item \textit{Type I} ($\tilde a_1\neq a_1$): not proportional to the
        ideal equations of motion; they change the physical meaning of the
        hydrodynamic variables \cite{Kovtun:2019hdm}.
          \item \textit{Type II} ($\tilde a_1=a_1$): proportional to the ideal
        equations of motion; they preserve the Landau-frame meaning.
\end{itemize}
Restricting to type-II frames is a central choice of this work.
As a result, neither the initial data nor the results need to be
transformed between frames: $\epsilon$ and $u^\mu$ can be taken directly
from, and used in, the Landau-frame descriptions standard in QGP
phenomenology.


Field redefinitions~\eqref{Field_redefinition_non_conformal2} modify the principal part of the equations, i.e., the terms with the highest derivatives, and thus affect their well-posedness. For type-II frames, the relativistic Navier--Stokes equations have a locally well-posed initial value problem and exhibit no faster-than-light propagation if
\begin{equation}
  a_1 \geq \frac{4}{3c_s^2}\,, \qquad
  a_2 \geq \frac{a_1}{\frac{3}{4}(1-c_s^2)^2 a_1-c_s^2}\,.
  \label{wellposedness_conditions0}
\end{equation}
For $c_s^2<1/2$, these conditions are equivalent to the general well-posedness and causality conditions of Bemfica, Disconzi and
Noronha~\cite{Bemfica:2019knx,Bemfica:2020gcl} restricted to type-II
frames: three of the latter are automatically satisfied, and the remaining
two are precisely Eq.~\eqref{wellposedness_conditions0} (see Appendix~A).
We thereby identify simple and compact conditions, convenient for phenomenological applications.
The terms $\mathcal{A}$ and $\mathcal{Q}^\mu$ thus act as regulators: they leave the first-order physics unchanged and, when
\eqref{wellposedness_conditions0} holds, restore well-posedness and causality, which are absent in the Landau frame, $a_1=a_2=0$.

Since Eq.~\eqref{wellposedness_conditions0} depends on the state through
$c_s^2(\epsilon)$, a key question is whether it can be satisfied for every
state. We show by explicit construction that this is possible, exploiting
the functional freedom in $\{a_1(\epsilon),a_2(\epsilon)\}$:
\begin{equation}
  \{a_1,a_2\}=\left\{\frac{25}{12\,c_s^2},\,
  \frac{100}{75-150\,c_s^2+27\,c_s^4}\right\},
  \label{sharply_causal_frame1}
\end{equation}
which satisfies Eq.~\eqref{wellposedness_conditions0} for any
$0<c_s^2<5/9$, a range that includes the QCD equation of state, and
saturates its second inequality (see Appendix~A for the characteristic speeds). Consequently, any evolution in this frame
is well-posed and causal by construction, for any initial data and at all
times, with no need to monitor causality during the simulation. 
A frame with the same $a_1$ was subsequently considered in Ref.~\cite{Patel:2026dba}.

Finally, for comparison we consider a theory of the MIS class, in the Landau frame, with stress tensor
\begin{equation}
T^{\mu\nu}=\epsilon u^\mu u^\nu+(p+\Pi)\Delta^{\mu\nu}+\pi^{\mu\nu}\,,
\label{MIS_stress_tensor}
\end{equation}
where the bulk and shear viscous stresses, $\Pi$ and $\pi^{\mu\nu}$, obey
\begin{subequations}
\begin{align}
\tau_\Pi\dot{\Pi}+\Pi &= -\zeta\nabla\cdot u\,,\\
\tau_\pi\Delta^{\mu\nu}{}_{\alpha\beta}\dot{\pi}^{\alpha\beta}+\pi^{\mu\nu} &= -\eta\sigma^{\mu\nu}\,,
\end{align}
\label{MIS_equations}
\end{subequations}
with $\tau_\Pi$ and $\tau_\pi$ their relaxation times and $\Delta^{\mu\nu}{}_{\alpha\beta}$ the transverse traceless projector. This is the theory of Ref.~\cite{Bemfica:2020xym}, 
whose causality conditions are known (see Appendix B), with all other second-order transport coefficients set to zero. In the EFT regime, these coefficients should not affect the first-order physics we study, and $\tau_\pi$ and $\tau_\Pi$ should act only as regulators.
A major advantage of relativistic Navier--Stokes over MIS theories is
precisely that, in the latter, the causality conditions depend on the
dissipative stresses and must be monitored during the evolution, whereas in
frame~\eqref{sharply_causal_frame1} they hold by construction.

\noindent
\textbf{3. Numerical evolutions.}
To model central collisions, we assume boost invariance along the beam axis and axial symmetry in the transverse plane. We use polar coordinates in the transverse plane and Milne coordinates in the longitudinal direction, $\{\tau,r,\phi,\eta_s\}$, related to Cartesian coordinates by
\begin{align}
&\{t,x,y,z\}=\{\tau \cosh{\eta_s} ,r \sin{\phi}  , r \cos{\phi} ,\tau \sinh{\eta_s}\} \,. 
\label{Milne_coordinates}
\end{align}
The four-velocity is $u^{\mu}=\{\sqrt{1+(u^r)^2},u^r,0,0\} \,$. We choose $\{\epsilon,u^r\}$ as our evolving variables.
We use the equation of state from the 2014 HotQCD Collaboration \cite{HotQCD:2014kol} and the shear and bulk viscosities from the JETSCAPE Collaboration \cite{Parkkila:2021tqq}. 
We solve the equations with sixth-order finite differences and fourth-order time integration; details of the code, the parametrizations and tests are given in Appendix~C and Ref.~\cite{SM}.

For initial conditions, we consider a smooth TRENTo~\cite{Moreland:2014oya,TRENToGitHub}  profile with entropy density proportional to the geometric mean of the nuclear thickness functions, $s\propto\sqrt{T_A T_B}$.
For central collisions $T_A(r)=T_B(r)=T(r)$, where $T(r)$ is obtained from a Woods--Saxon nuclear density distribution
\begin{equation}
T(r)=\int_{-\infty}^{\infty}
\left(
1+e^{\frac{\sqrt{r^2+z^2}-R_0}{a_0}}
\right)^{-1} dz \,,
\end{equation}
where $(R_0,a_0)=(6.62,0.546)\,\mathrm{fm}$ are
the nuclear radius and skin width of Pb~\cite{DeVries:1987atn}. 
The initial energy density is obtained from the entropy
density using the equation of state,
\begin{subequations}
\begin{align}
s(\epsilon)|_{\tau_0}
&=s(\epsilon_0)\frac{T(r)}{T(0)}\,, \label{initial_energy}
\\
u^r|_{\tau_0}&=0\,,
\end{align}
\label{initial_data_2}
\end{subequations}
where $\tau_0$ is the initial proper time, $\epsilon_0:=\epsilon|_{\tau_0,r=0}$ is the
initial energy density at the center, and we assume vanishing initial transverse velocity.

Since the equations are second order in time, the initial data must also include the first time derivatives of the evolved variables. 
We fix them by requiring that the ideal equations of motion~\eqref{EOMs_ideal_hydrodynamics} hold at $\tau_0$, which gives
\begin{subequations}
	\begin{align}
		\partial_{\tau} \epsilon|_{\tau_0}&= -\frac{p(\epsilon |_{\tau_0})+\epsilon |_{\tau_0}}{\tau_0}\,, \label{Initial_data_boost_c}\\
		\partial_{\tau}	u^r|_{\tau_0}&=-\frac{\partial_{\epsilon}p(\epsilon |_{\tau_0})\, \partial_r{\epsilon |_{\tau_0}} }{p(\epsilon |_{\tau_0})+\epsilon |_{\tau_0}}  \,. \label{Initial_data_boost_d}
	\end{align}
	\label{Initial_data_boost_inv1}
\end{subequations}
In the absence of a pre-equilibrium model that determines the initial time
derivatives, such as IP-Glasma~\cite{Schenke:2018fci},
a system described by
hydrodynamics should satisfy Eqs.~\eqref{Initial_data_boost_inv1} up to
first-order corrections, which are small within the EFT regime.
This prescription, which proved reliable for a conformal fluid in
Ref.~\cite{Bea:2023rru}, is extended here to a realistic heavy-ion setting.
With this prescription, $\mathcal{A}$ and $\mathcal{Q}^\mu$ vanish at $\tau_0$.

\begin{figure}[t]
          \includegraphics[width=0.5295\columnwidth]{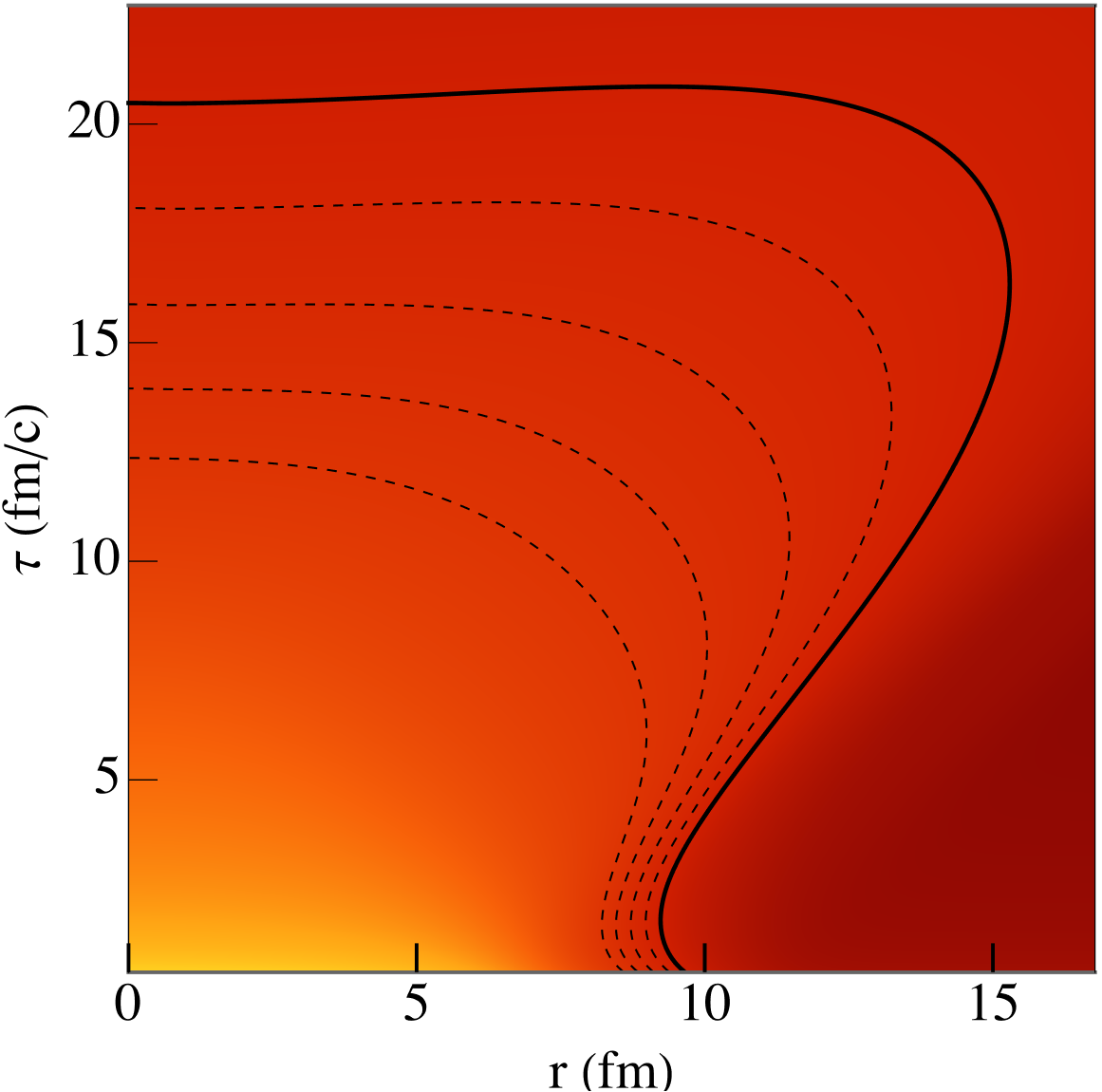}\hspace{-0.04cm}
          \put (-84,128) {\small  Navier--Stokes }
                    \put (49,128) {\small  MIS }
      \includegraphics[width=0.475\columnwidth]{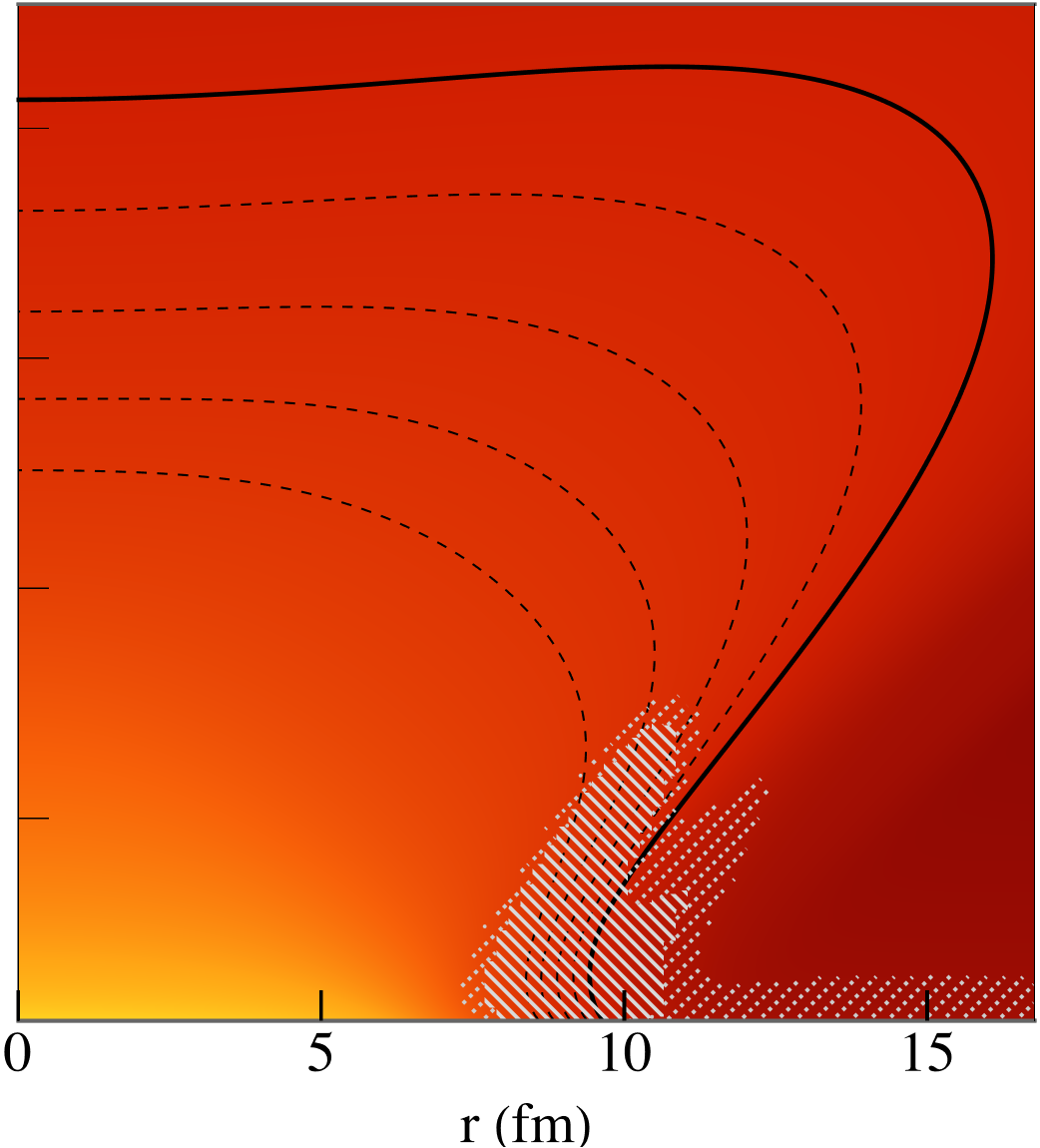}
            \includegraphics[width=\columnwidth]{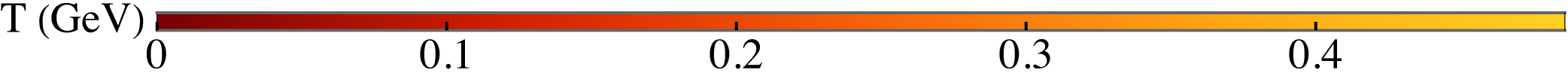}
  \caption{Temperature as a function of proper time and radius  for relativistic
Navier--Stokes (left) and MIS (right). Solid lines: freeze-out surface;
dashed lines: isotherms $T=0.13$--$0.16$~GeV. Hatched (dotted) region: the
necessary (sufficient) causality conditions of
Ref.~\cite{Bemfica:2020xym} are violated (not satisfied). The dark region at large $r$ is the low-density regulator.}
  \label{fig:RNS_MIS_comparison}
\end{figure}

We evolve the relativistic Navier--Stokes equations in the
frame~\eqref{sharply_causal_frame1} with initial
data~\eqref{initial_data_2} and~\eqref{Initial_data_boost_inv1} for
different values of $\{\tau_0,\epsilon_0\}$, defining the freeze-out surface
as the isotherm $T_f=0.12$~GeV. Figure~\ref{fig:RNS_MIS_comparison} (left)
shows the evolution for $\tau_0=0.6$~fm/$c$ and
$\epsilon_0=110$~GeV/fm$^3$.

For comparison, Fig.~\ref{fig:RNS_MIS_comparison} (right) shows an evolution
of the MIS theory~\eqref{MIS_stress_tensor} with the same equation of state,
viscosities and initial data~\eqref{initial_data_2}, and the relaxation
times of MUSIC~\cite{Schenke:2010nt,Schenke:2010rr,Paquet:2015lta},
Eq.~\eqref{taus}. 
The viscous stresses are initialized at their first-order values, 
$\pi^{\mu\nu}(\tau_0)=-\eta\sigma^{\mu\nu}$ and
$\Pi(\tau_0)=-\zeta\,\nabla\ydot u$, so that both theories start from the
 same stress tensor. 
 The necessary causality conditions of Ref.~\cite{Bemfica:2020xym} are
violated in the hatched region, at early times and extending inside the
freeze-out surface, consistent with previous
studies~\cite{Plumberg:2021bme,Chiu:2021muk,ExTrEMe:2023nhy,Domingues:2024pom}.

\begin{figure}[thbp]
\centering
		{\includegraphics[width=\columnwidth]{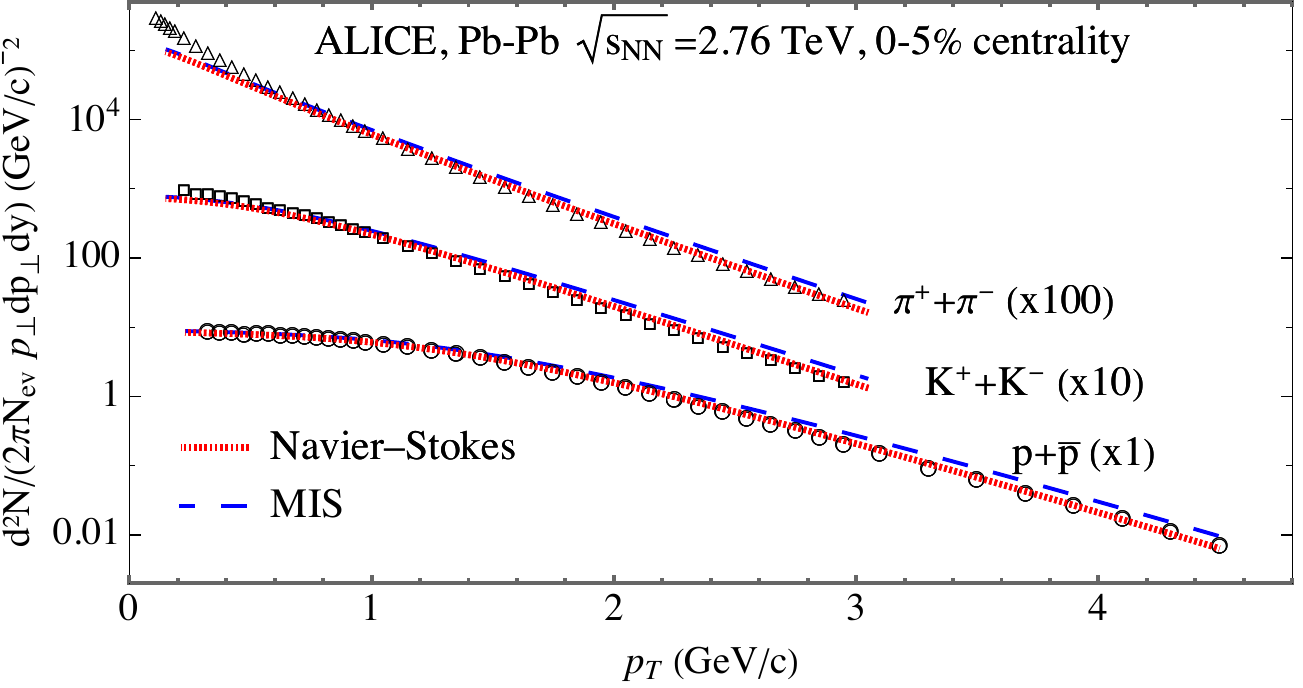}}  
\caption{Transverse-momentum spectra of hadrons in 0--5\% central Pb--Pb
collisions at $\sqrt{s_{NN}}=2.76$~TeV measured by
ALICE~\cite{ALICE:2013mez}, compared with those from the relativistic
Navier--Stokes (red dotted) and MIS (blue dashed) evolutions.}
	\label{Description_experimental_data_1}
\end{figure}

\noindent
\textbf{4. Comparison with data.} We compare with the transverse-momentum spectra of pions, kaons and protons
in 0--5\% central Pb--Pb collisions at $\sqrt{s_{NN}}=2.76$~TeV measured by
ALICE~\cite{ALICE:2013mez,HEPData:ALICE:2013}. 
At the freeze-out surface, the fluid is converted into hadrons with thermal
distributions boosted by the local flow, via the Cooper--Frye
formula~\cite{Cooper:1974mv,SM}.
Scanning $\{\tau_0,\epsilon_0,T_f\}$, we find good agreement
for
\begin{equation}
  \tau_\mathrm{0}\!=\!0.6\,\mathrm{fm}/c,\
  \epsilon_\mathrm{0}\!=\!110\,\mathrm{GeV/fm^3},\
  T_\mathrm{f}\!=\!0.12\,\mathrm{GeV},
  \label{best_fit_params}
\end{equation}
The resulting spectra are shown in Fig.~\ref{Description_experimental_data_1};
the evolution in Fig.~\ref{fig:RNS_MIS_comparison} (left) corresponds
precisely to these parameters.
The same parameters describe the three species, including the flattening of the
spectra of heavier hadrons at low $p_T$, a distinctive signature of radial
flow~\cite{SM}. The mean radial velocity at freeze-out,
$\langle v_r\rangle\simeq0.67$, is close to the ALICE blast-wave value
$\langle\beta_T\rangle=0.65\pm0.02$~\cite{ALICE:2013mez}. 

The MIS evolution of Fig.~\ref{fig:RNS_MIS_comparison} (right) 
leads to the blue dashed spectra in Fig.~\ref{Description_experimental_data_1},
which also describe the data, with slight deviations: the MIS spectra are
slightly harder, consistent with its wider freeze-out surface.
One source of these deviations is
that second-order effects in MIS are not small, so that the evolution lies
outside the EFT regime at early times, as we show in Sec.~5. Moreover, the
MIS evolution violates the necessary causality conditions at early times,
which may also contribute to these deviations.
Note that the parameters~\eqref{best_fit_params} were fitted with
Navier--Stokes and then used for MIS, so these deviations do not indicate
that one theory describes the data better: with refitted parameters, MIS
could describe the data equally well. The choice between the two theories
must therefore rest on theoretical grounds. Within the EFT regime, both
should give equivalent descriptions up to higher-order terms, so sizable
differences signal a departure from this regime; moreover, a theory that
violates causality or well-posedness cannot provide a reliable description,
regardless of its agreement with data. As we show in Sec.~5, relativistic
Navier--Stokes satisfies both criteria in this setup, whereas MIS does not
at early times.

Since our approach does not include chemical freeze-out, each species is
normalized by a constant factor, common to the Navier--Stokes and MIS
evolutions, so the comparison probes the spectral shapes; the pion excess
at low $p_T$ reflects resonance decays, not included here.

\noindent
\textbf{5. Effective field theory analysis.} To assess whether the solution in Fig.~\ref{fig:RNS_MIS_comparison} (left)
lies within the EFT regime, we compare, at each time, the $L_1$ norms (sums
of absolute values over grid points) of the viscous and ideal contributions
to $T^{rr}$ in Eq.~\eqref{stress_tensor_generic_frame},
$-\eta\sigma^{rr}-\zeta\,\nabla\ydot u\,\Delta^{rr}$ and
$\epsilon\,u^ru^r+p\,\Delta^{rr}$~\cite{SM}. Their ratio is $32\%$ at
$\tau_0=0.6$~fm/$c$, so the system starts marginally within the EFT regime,
and decreases to $20\%$, $10\%$ and $3\%$ at $\tau=1.2$, $2.8$ and
$6.0$~fm/$c$, respectively, so the system enters progressively deeper into
the EFT regime; the terms $\mathcal{A}$ and $\mathcal{Q}^\mu$ remain smaller than the viscous ones (see Ref.~\cite{SM}).

For the MIS evolution in Fig.~\ref{fig:RNS_MIS_comparison} (right), the
ratio coincides at $\tau_0$ by construction, but then grows to $38\%$ at
$\tau=1.2$~fm/$c$, nearly twice the Navier--Stokes value, and remains larger,
$16\%$ and $5\%$ at $\tau=2.8$ and $6.0$~fm/$c$. Repeating the evolution with
$\tau_\pi\to\tau_\pi/2$ reduces the ratio to $28\%$, $12\%$ and $4\%$, closer
to the Navier--Stokes values. This identifies the origin of the large viscous
contribution: since $\tau_\pi$ is comparable to $\tau_0$, $\pi^{\mu\nu}$
cannot follow the rapid decay of its Navier--Stokes value and, at early times,
decays more slowly than the ideal pressure~\cite{SM}. These second-order
effects are therefore not small in this regime.

As a self-consistency check of the MIS evolution, we compare the evolved
dissipative contribution, $\pi^{rr}+\Pi\,\Delta^{rr}$, with its first-order
value, $-\eta\sigma^{rr}-\zeta\,\nabla\ydot u\,\Delta^{rr}$, computed from the
evolved $\epsilon$ and $u^\mu$. Within the regime of
validity of the gradient expansion, both should agree up to higher-order
corrections. Instead, the former is about twice the latter along the evolution, showing that
second-order effects, here dominated by the relaxation lag of $\pi^{\mu\nu}$,
are comparable to first-order ones.

Returning to relativistic Navier--Stokes, a further consistency test
follows from the freedom in the choice of hydrodynamic frame. Different
frames differ only by terms beyond first order, so within the EFT regime the first order
physics must not depend on this choice. 
We therefore repeat the evolution of Fig.~\ref{fig:RNS_MIS_comparison} (left) in a second frame, with $a_1$ reduced by a factor $16/25$,
\begin{equation}
\{a_1,a_2\}=\Big\{\frac{4}{3c_s^2},\,\frac{4}{3-6c_s^2}\Big\}\,,
\label{sharply_causal_frame2}
\end{equation}
which also satisfies conditions~\eqref{wellposedness_conditions0}, and perform a third evolution 
using the ideal hydrodynamics equations and the same initial data. At each $\tau$ we
compare the $L_1$ norm of the frame difference,
$T^{rr}_{\mathrm{NS(8)}}-T^{rr}_{\mathrm{NS(16)}}$, with that of the viscous
correction, $T^{rr}_{\mathrm{NS}(8)}-T^{rr}_{\mathrm{ideal}}$, where
$\mathrm{NS(8),(16)}$ denote the evolutions in
frames~\eqref{sharply_causal_frame1} and~\eqref{sharply_causal_frame2};
this corresponds to Criterion C of \cite{Bea:2023rru}.
Their ratio remains about $1\%$ at all times: the frame dependence is two
orders of magnitude smaller than the effect of viscosity itself, and we conclude that the
relativistic Navier--Stokes prediction is robust under the choice of frame.

Finally, we compare the effect of using different frames with the second-order effects in MIS and the ideal evolution. 
 The difference
$T^{rr}_{\mathrm{NS(8)}}-T^{rr}_{\mathrm{MIS}}$ reaches ${\sim}120\%$ of the viscous
correction, $T^{rr}_{\mathrm{NS(8)}}-T^{rr}_{\mathrm{ideal}}$, and halving
$\tau_\pi$ reduces it to ${\sim}60\%$ (see Ref.~\cite{SM} for the time dependence of these ratios and the
spatial profiles of $T^{rr}$). 
This further confirms that the second-order effects the MIS evolutions are far larger than the frame
dependence of relativistic Navier--Stokes, and that the MIS evolution lies outside the EFT regime.

A natural solution would be to reduce $\tau_\pi$, bringing the MIS evolution
closer to the EFT regime. However, this considerably enlarges the regions
where the necessary causality conditions are violated, Fig.~\ref{MIS_difference_espectra_1} (top right), so
that, in this MIS setup, causality and small second-order corrections cannot be
achieved simultaneously, revealing a tension between them.


Larger deviations of minimal Israel--Stewart theory from ideal hydrodynamics
at early times were also observed in Ref.~\cite{Patel:2026dba}; the analysis
above shows that they are second-order effects, signaling that the evolution
lies outside the EFT regime.

\noindent
\textbf{6. Discussion.} We have shown that relativistic Navier--Stokes hydrodynamics in the BDNK
formulation can describe the radial flow measured at the LHC, providing a
proof of principle for its viability in heavy-ion phenomenology.


On the theoretical side, we have presented the relativistic Navier--Stokes BDNK
equations in a general frame via a field redefinition of the standard
Landau-frame stress tensor, which makes their EFT interpretation explicit.
We identified a class of frames, which we termed type~II, in which the
hydrodynamic variables retain their Landau-frame interpretation, since the
additional terms $\mathcal{A}$ and $\mathcal{Q}^\mu$ in \eqref{stress_tensor_generic_frame} are proportional to the
ideal equations of motion and hence of second order on shell. For these
frames, we showed that the well-posedness and causality conditions of
Bemfica, Disconzi and Noronha~\cite{Bemfica:2019knx,Bemfica:2020gcl} reduce
to the simple conditions~\eqref{wellposedness_conditions0}.
This leads to a key message for phenomenological applications: adding
these terms leaves the first-order physics unchanged, regularizes the
relativistic Navier--Stokes equations, and preserves the Landau-frame
interpretation of the hydrodynamic variables.

We performed boost-invariant, axially symmetric simulations with a QCD
equation of state and transport coefficients. For the initial conditions, we
employed a smooth TRENTo-motivated profile 
with vanishing initial transverse
velocity~\eqref{initial_data_2}. 
An important advantage of type-II frames for phenomenological applications
is that neither the initial data nor the final solution need to be
transformed to the Landau frame, since the hydrodynamic variables retain
their Landau-frame interpretation.
The initial time derivatives, required
since the equations are second order in time, were fixed by the ideal
hydrodynamic equations~\eqref{Initial_data_boost_inv1}. 
In the non-conformal case, where conditions~\eqref{wellposedness_conditions0} depend on the local state, we constructed a frame~\eqref{sharply_causal_frame1} that satisfies them for every state, ensuring causality and well-posedness throughout any evolution. 
The evolution in Fig.~\ref{fig:RNS_MIS_comparison} (left) describes well the hadron spectra measured by ALICE, Fig.~\ref{Description_experimental_data_1}.
In this evolution, the ratio of viscous to ideal terms is initially $32\%$ and decreases rapidly over time, indicating that the system enters the EFT
regime. 
Moreover, the difference between evolutions in two distinct frames is much smaller---about $1\%$---than their difference from an ideal
hydrodynamic evolution, showing that the predictions of relativistic Navier--Stokes are robust under the choice of frame.

For comparison with standard frameworks, we performed simulations of a
theory of the MIS class~\eqref{MIS_stress_tensor} with the same equation of
state, viscosities and initial data, initializing the dissipative stresses
at their Navier--Stokes values and using the MUSIC relaxation times, and
found causality violations, Fig.~\ref{fig:RNS_MIS_comparison} (right).
Within the MIS evolution
itself, the dissipative stresses are about twice their first-order values,
computed from the evolved energy density and velocity, showing that
second-order effects are as large as first-order ones.
Halving $\tau_\pi$, which should act only as a regulator, changes the MIS
results substantially, confirming that these effects are of second order. 
These effects also reach the observables: with the same parameters, the MIS
spectra deviate from the data (Fig.~\ref{Description_experimental_data_1}),
and halving $\tau_\pi$ brings them closer to the relativistic Navier--Stokes
result (Fig.~\ref{MIS_difference_espectra_1}).
Since viscosities are extracted by
fitting MIS-based evolutions to data, such effects may bias the extracted
values, particularly at the high temperatures probed early in the evolution.
The difference between the MIS and relativistic Navier--Stokes evolutions
reaches about $120\%$ of the viscous correction, i.e., of the difference
between relativistic Navier--Stokes and ideal hydrodynamics, and depends
strongly on the relaxation time, whereas the frame dependence of relativistic
Navier--Stokes is only about $1\%$ of the same quantity.
A natural remedy would be to reduce $\tau_\pi$, but this considerably
enlarges the regions where causality is violated, revealing a tension
between small second-order effects and causality within this MIS setup.
This is an important qualitative lesson of our study, and raises the
question of whether such a tension is generic in MIS-based simulations.

This work lays the foundation for more realistic relativistic Navier--Stokes simulations of heavy-ion collisions, including event-by-event simulations
and small systems, where gradients are larger, offering a promising alternative to MIS-based descriptions.

\begin{acknowledgments}
\textbf{Acknowledgments.} I thank Jorge Casalderrey-Solana, Pau Figueras, Raimon Luna, David Mateos, Alexandre Serantes and Carlos V\'azquez Sierra for very useful discussions.
I acknowledge support by the Beatriu de Pin\'os postdoctoral program under the Ministry
of Research and Universities of the Government of Catalonia (2022 BP 00225), by a Maria Zambrano postdoctoral fellowship from the University of Barcelona, by the “Unit of Excellence MdM 2020-2023” award to the Institute of Cosmos Sciences (CEX2019-000918-M) funded by MCIN/AEI/ 10.13039/501100011033/FEDER, UE, and 
grants PID2019-105614GB-C21, PID2019-105614GB-C22, PID2022-136224NB-C21, PID2022-136224NB-C22 and 2021-SGR-872.
This work was also supported by grant ED431B 2024/42 (Consellería de Cultura, Educación, Formación Profesional y Universidades, Xunta de Galicia).
This publication is part of the R\&D\&i projects PID2025-172448NB-I00 and
PID2025-169343NB-C21, funded by MICIU/AEI/10.13039/501100011033.
\end{acknowledgments}

\bibliography{refs_BDNK}


\clearpage

\onecolumngrid
\begin{center}
    {\large\bfseries End Matter}
\end{center}
\twocolumngrid

\appendix
\setcounter{secnumdepth}{1}





\noindent
\textbf{Appendix A. Type-II frames.}
\textit{Well-posedness and causality conditions.}
Here we show that, for type-II frames and $c_s^2<1/2$, the
conditions~\eqref{wellposedness_conditions0} are equivalent to the general
well-posedness and causality conditions of Ref.~\cite{Bemfica:2019knx}.
The stress tensor~\eqref{stress_tensor_generic_frame} is of the form
considered there, with $\chi_1=a_1(\eta+\tfrac{3}{4}\zeta)$,
$\chi_2=\tilde a_1(\eta+\tfrac{3}{4}\zeta)$, $\chi_3=c_s^2\chi_1$,
$\chi_4=c_s^2\chi_2-\zeta$ and $\lambda=a_2(\eta+\tfrac{3}{4}\zeta)$. For type-II frames, $\chi_1=\chi_2$, and conditions~(7b)
and~(7c) of Ref.~\cite{Bemfica:2019knx} become precisely
Eq.~\eqref{wellposedness_conditions0}; the requirement $c_s^2<1/2$ ensures
that the denominator in Eq.~\eqref{wellposedness_conditions0} is positive.
In turn, Eq.~\eqref{wellposedness_conditions0} implies $a_1,a_2>0$ and
$a_2>4/[3(1-c_s^2)^2]$, from which the remaining conditions follow.
Condition~(6) of Ref.~\cite{Bemfica:2019knx} becomes
\begin{equation}
  16\,a_1\big(\eta+\tfrac{3}{4}\zeta\big)^4
  \left(a_1+3c_s^2a_1a_2+3c_s^2a_2^2\right)\geq 0\,,
\end{equation}
which holds for $a_1,a_2>0$, while conditions~(7a) and~(7d) become
$a_2(\eta+\tfrac{3}{4}\zeta)\geq\eta$ and $a_2\geq 2/[3(1-c_s^2)]$, which
follow from $a_2>4/[3(1-c_s^2)^2]$. This completes the proof.

In the conformal case 
Eq.~\eqref{wellposedness_conditions0} becomes
$a_1\geq4$ and $a_2\geq3a_1/(a_1-1)$, which are the conditions of
Ref.~\cite{Bemfica:2017wps}.

\textit{Characteristic speeds.}
The characteristic speeds of a system of partial differential equations are the speeds at which
information propagates. For frames satisfying Eq.~\eqref{wellposedness_conditions0},
they do not exceed the speed of
light~\cite{Bemfica:2019knx,Bemfica:2020gcl}. 
Under our symmetry assumptions, the characteristic speeds are $\pm c_+$
and $\pm c_-$, with
\begin{equation}
  c_\pm^2=c_s^2+\frac{2\big(a_1\pm\sqrt{a_1\left[a_1+3a_2(a_1+a_2)c_s^2\right]}\big)}{3a_1a_2}\,,
  \label{characteristic_velocities}
\end{equation}
which for the frame~\eqref{sharply_causal_frame1} reduce to $c_+=1$ and
$c_-=3c_s^2/5$.
Since this frame saturates the second
inequality in~\eqref{wellposedness_conditions0}, its largest
characteristic speed equals the speed of light for every state, i.e., it
is sharply causal~\cite{Freistuhler:2021lla}. In the conformal limit, it
reduces to the sharply causal frame of Ref.~\cite{Pandya:2021ief}, for
which shocks of arbitrary strength were found to be smoothly resolved.
This property is essential here: a strong shock forms between the
fireball and the low-density regulator region, and in frames that are
not sharply causal, shocks beyond a critical strength admit no smooth
solution~\cite{Pandya:2021ief,Bea:2023rru}.


\noindent
\textbf{Appendix B. MIS causality conditions.} For the relaxation times we use the values implemented in MUSIC~\cite{Schenke:2010nt,Schenke:2010rr,Paquet:2015lta},
\begin{equation}
  \tau_\pi = \frac{5\,\eta}{\epsilon+p}\,, \qquad
  \tau_\Pi = \frac{\zeta}{15\,(1/3-c_s^2)^2\,(\epsilon+p)}\,.
  \label{taus}
\end{equation}
With these values and $0<c_s^2<1/3$, the necessary ($\alpha=0$) and
sufficient ($\alpha=1$) causality conditions of
Ref.~\cite{Bemfica:2020xym} reduce to
\begin{equation}
  \frac{\Pi+\Lambda_1}{\epsilon+p}
  -\alpha\,\frac{c_s^2\,(\Lambda_3-\Lambda_1)}{(1-c_s^2)(\epsilon+p)}
  \;\geq\;
  \frac{(2-15\,c_s^2)(7-15\,c_s^2)}{15\,(1-c_s^2)}\,,
  \label{eq:necessary}
\end{equation}
where $\Lambda_1$ and $\Lambda_3$ are the smallest and largest eigenvalues of $\pi^\mu{}_\nu$ in the subspace orthogonal to $u^\mu$.
Violation of the necessary conditions implies superluminal characteristic speeds or loss of hyperbolicity. In
Fig.~\ref{fig:RNS_MIS_comparison} (right), the necessary condition is violated in the hatched region, and only the sufficient one fails in the
dotted region; elsewhere, the sufficient conditions are satisfied.

Note that, although the second-order transport coefficients set to zero in
Eqs.~(10) do not affect the first-order physics in the EFT regime, they do
enter the causality conditions~\cite{Bemfica:2020xym} and can therefore
modify them.

\noindent
\textbf{Appendix C. Numerical code and tests.}
Our numerical code is implemented in C++ and Mathematica.
Under our symmetry assumptions and coordinate choice, the relativistic Navier--Stokes equations reduce to 1+1 dimensional partial differential equations in variables $\tau$ and $r$. 
Since they are second order in time, we solve them explicitly for $\partial_\tau^2\epsilon$ and $\partial_\tau^2u^r$, and reduce them to a first-order system by introducing $\partial_\tau\epsilon$ and $\partial_\tau u^r$ as additional variables. 
We use a uniform grid in $r$, sixth-order finite differences, and the fourth-order Adams--Bashforth method, initialized with fourth-order Runge--Kutta, with Courant factor $\Delta\tau/\Delta r=0.1$ . We include Kreiss-Oliger dissipation of order 6 with coefficient $\sigma_{KO}=0.4$. Details of these methods are given in the Supplemental Material.
The polar coordinates used in the transverse plane $\{r,\phi\}$ introduce a singularity at the origin. 
We included the singular point $r=0$ in the grid, and in the equations of motion we used L'H\^opital's rule obtaining a non-singular expression. 
Details of these methods 
are given in Ref.~\cite{SM}.

\begin{figure}[thbp]
\centering
	{\includegraphics[width=\columnwidth,height=0.45\columnwidth]{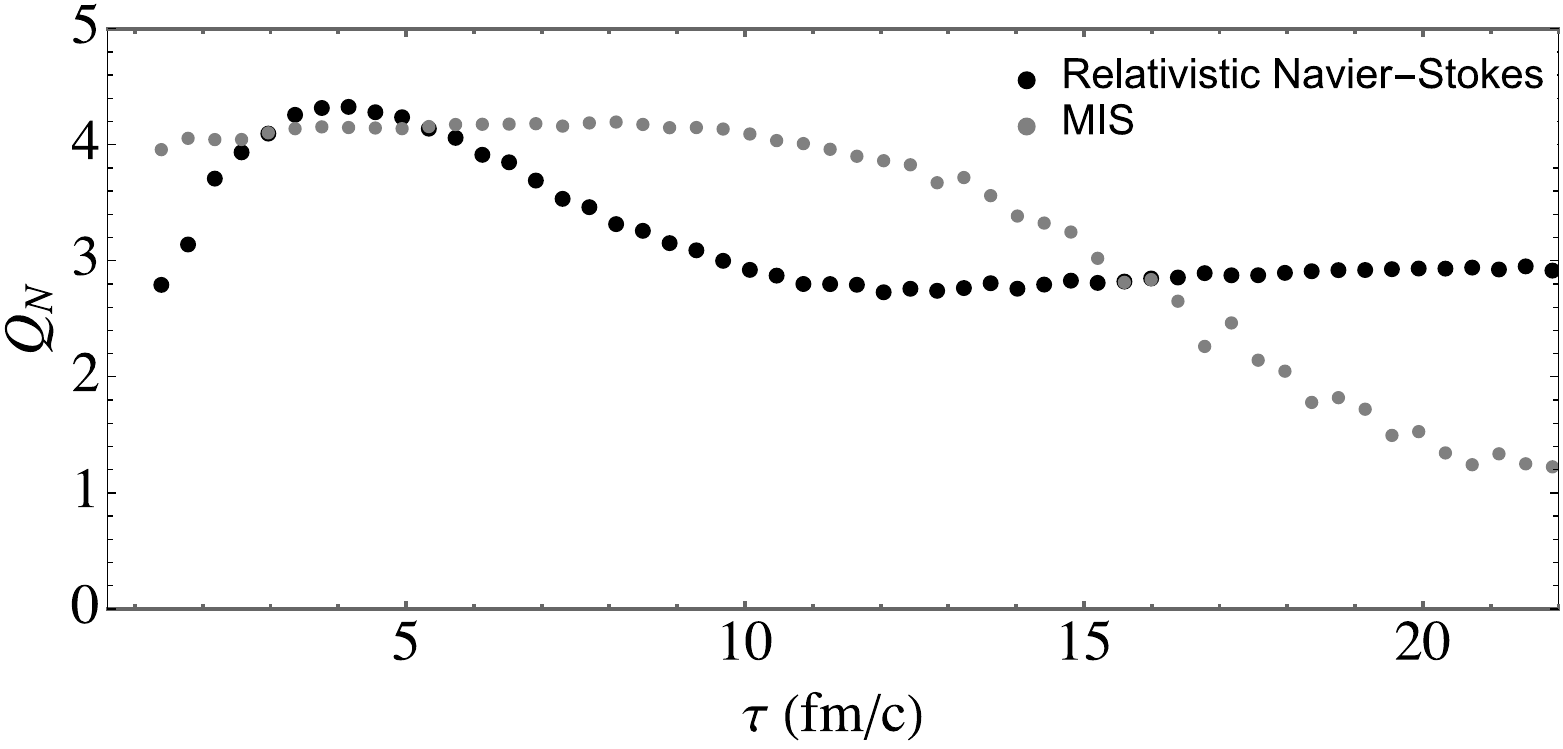}} 
	\caption{ Convergence order $\mathcal{Q}_N$ for the
evolutions of Fig.~\ref{fig:RNS_MIS_comparison}. Black: relativistic Navier--Stokes; grey: MIS.
	}
	\label{Convergence_tests_1}
\end{figure}

In Fig.~\ref{Convergence_tests_1}, we show the convergence order, defined in Ref.~\cite{SM}, for the evolution of Fig.~\ref{fig:RNS_MIS_comparison} (left), with resolutions $N=\{200,400,800\}$. According to our numerical scheme, the expected convergence order is 4. However, we find slightly lower values, closer to 3, in some regions of the time domain, which we attribute to the front between the fireball and the low-density regulator region, outside the freeze-out surface, where gradients are steep. 
%
As is standard, we include this regulator, sometimes called a ``corona'', by adding a small constant energy density, $\epsilon_\mathrm{reg}=0.0007\,\mathrm{GeV/fm^3}$, to the initial profile~\eqref{initial_energy}. 
We have verified that multiplying or dividing $\epsilon_\mathrm{reg}$ by a factor of two leaves the results unchanged: the freeze-out surface and
the spectra are indistinguishable at the scale of the figures.

For the MIS evolutions, we use the same numerical scheme. Since the MIS equations are already first order in time, no reduction is needed. A
stable evolution requires a smaller Courant factor, $\Delta\tau/\Delta r=0.02$. The convergence test, shown in
Fig.~\ref{Convergence_tests_1}, gives a convergence order that decreases at late times, which originates in the low-density region outside the freeze-out surface. 

\begin{figure}[thbp]
\centering
	{\includegraphics[width=1\columnwidth,height=0.5\columnwidth]{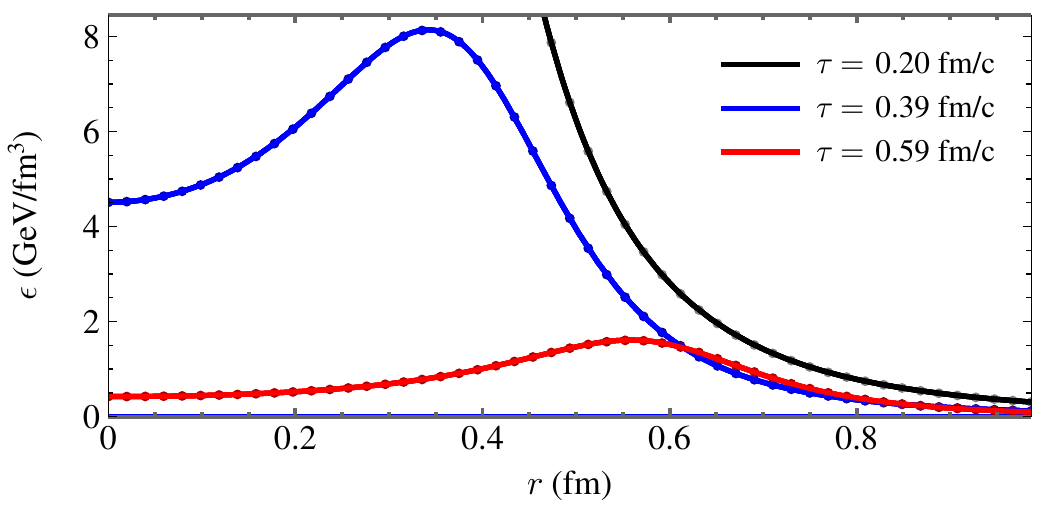}}
	{\includegraphics[width=1\columnwidth,height=0.5\columnwidth]{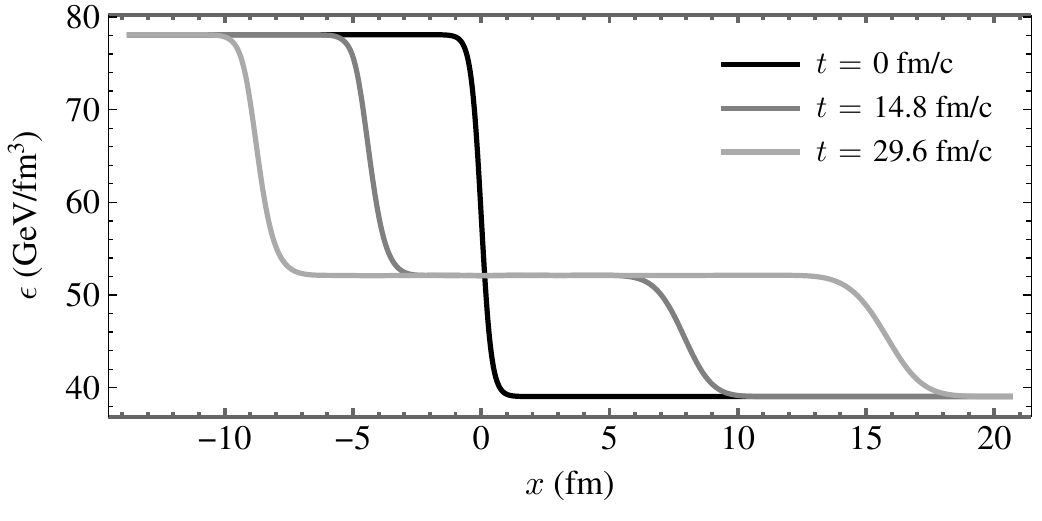}}
\caption{Top: Gubser flow test. Semi-analytic solution (dotted) and
numerical evolution (solid) at $\tau=0.20$, $0.39$ and $0.59\,\mathrm{fm}/c$;
the profile at $\tau=0.20\,\mathrm{fm}/c$ is the initial data. Bottom:
Riemann problem at $t=0$, $14.8$ and $29.6\,\mathrm{fm}/c$, showing a
rarefaction wave (left) and a smooth shock wave (right).}
\label{fig:tests}
	\label{Further_tests}
\end{figure}

\textit{Further tests: Gubser flow and Riemann problem.}
We further validate the code with two standard benchmarks, shown in
Fig.~\ref{fig:tests}. The first is Gubser flow~\cite{Gubser:2010ze,Gubser:2010ui},
a boost-invariant, axially symmetric flow of a conformal fluid, for which
the equations of motion reduce to ordinary differential equations in de
Sitter coordinates. We solve these equations, map the solution to
$(\tau,r)$ through the corresponding Weyl rescaling, and use the energy
density, the velocity and their time derivatives at
$\tau=0.20\,\mathrm{fm}/c$ as initial data for our code, with a conformal
equation of state and the shear viscosity and frame of
Ref.~\cite{Bea:2023rru}. The numerical evolution agrees with the
semi-analytic solution at later times, Fig.~\ref{fig:tests} (top).
The second is the Riemann problem, in which two homogeneous states are
initially separated by a sharp transition. We solve it with a Cartesian
version of our code, assuming homogeneity in two directions, with the
QCD equation of state and transport coefficients used in this work, in
frame~(8), and initial data at rest interpolating between
$\epsilon=78$ and $39\,\mathrm{GeV/fm^3}$ through a hyperbolic tangent,
similar to those of Sec.~3.3 of Ref.~\cite{Bea:2023rru}. The solution
develops a rarefaction wave and a shock wave, which is smooth due to
viscosity, Fig.~\ref{fig:tests} (bottom), in agreement with the expected
structure of the solution~\cite{Rezzolla:2013dea}.

\noindent
\textbf{Appendix D. Frame and relaxation-time dependence of observables.}

\begin{figure}[thbp]
          \includegraphics[width=0.5295\columnwidth]{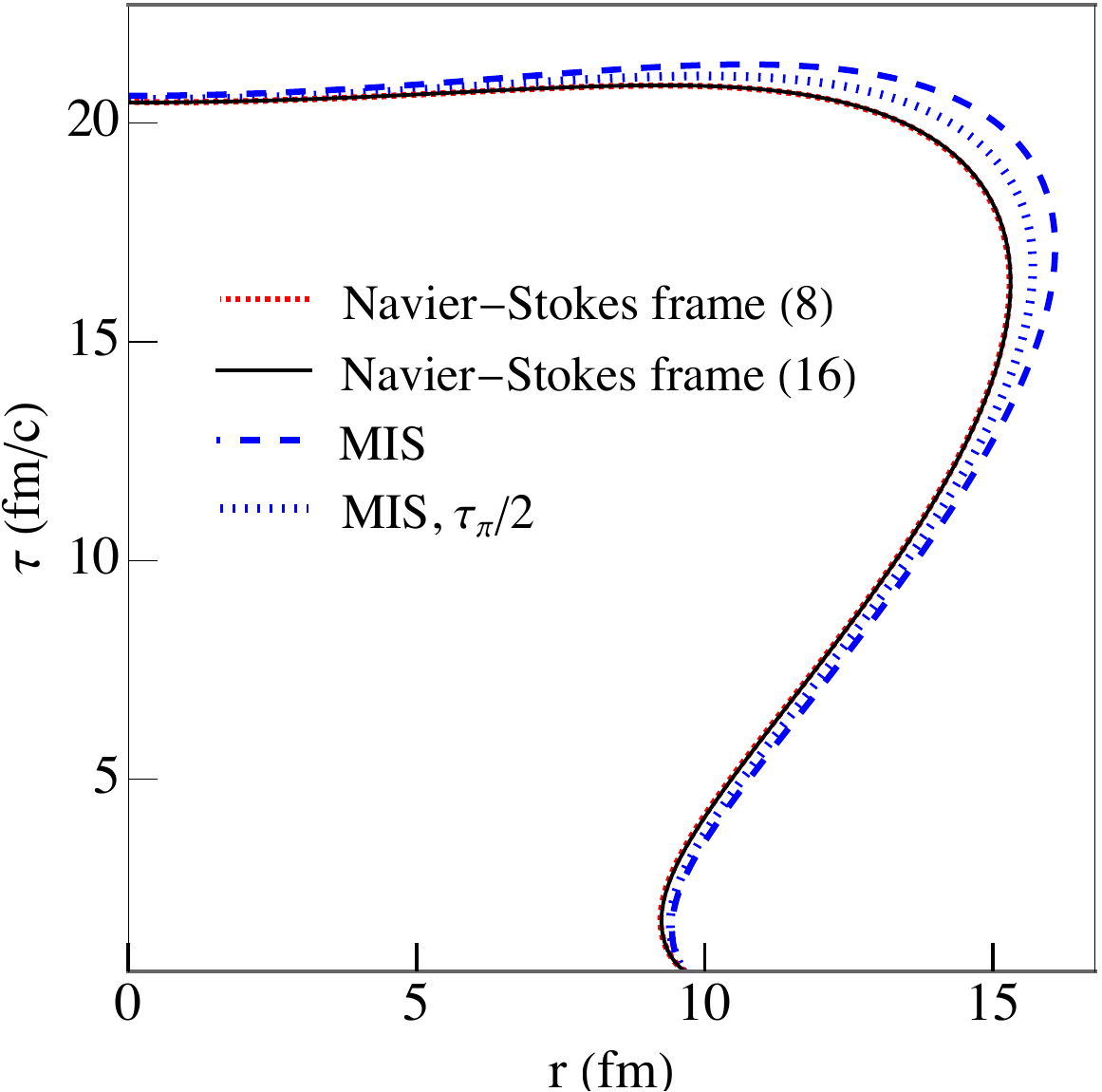}\hspace{-0.04cm}
	          \put (-90,128) {\small  Freezout surfaces }
                    \put (27,128) {\small  MIS with $\tau_{\pi}/2$}
      \includegraphics[width=0.475\columnwidth]{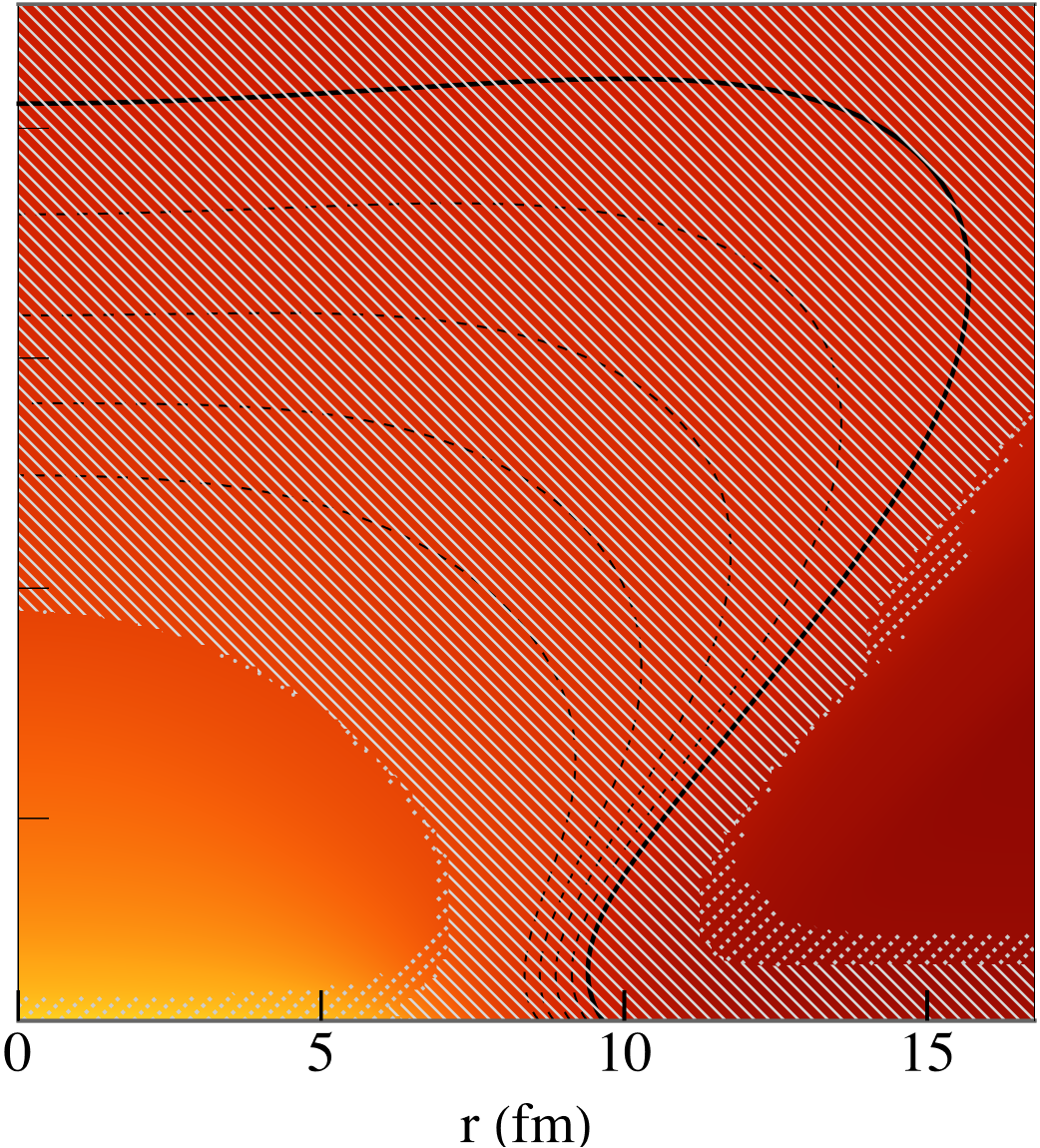}
	            \includegraphics[width=\columnwidth]{./Hydro/BAR_EV.pdf}\vspace{2mm}
	{\includegraphics[width=\columnwidth]{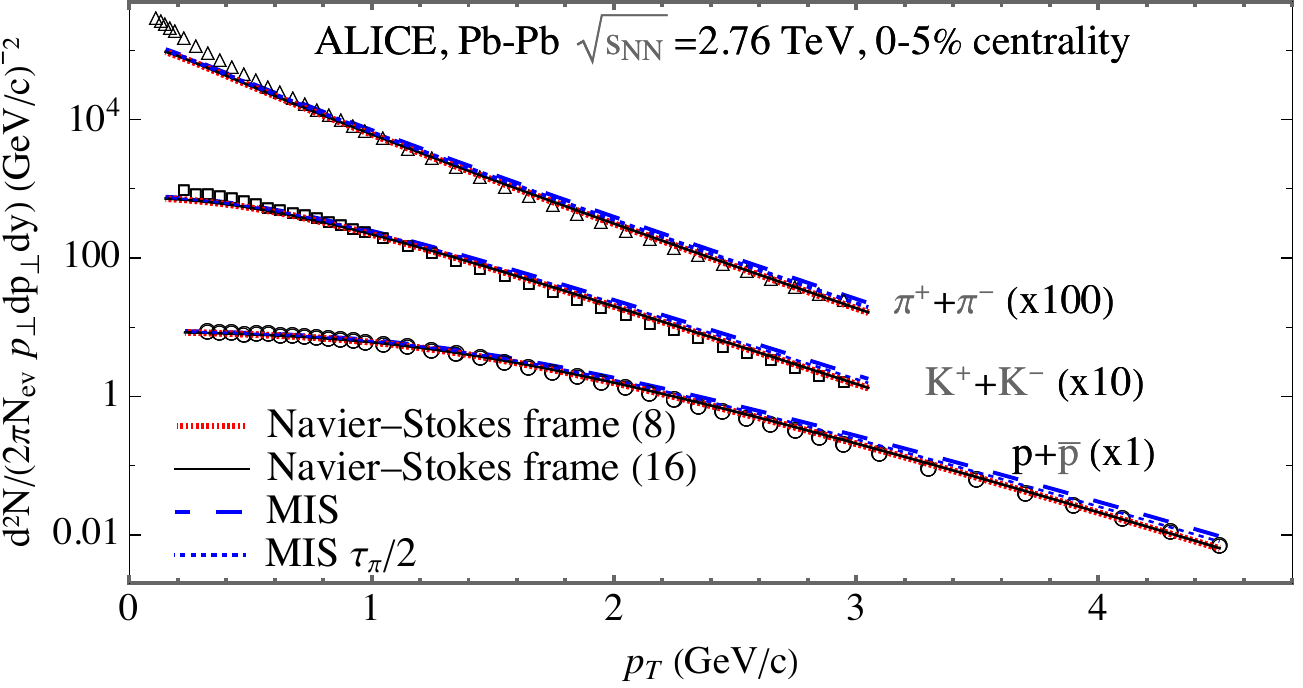}} 
\caption{Top left: freeze-out surfaces for relativistic Navier--Stokes in
frames~\eqref{sharply_causal_frame1} and~\eqref{sharply_causal_frame2}, and
for MIS with $\tau_\pi$ and $\tau_\pi/2$. Top right: as
Fig.~\ref{fig:RNS_MIS_comparison} (right), for MIS with $\tau_\pi/2$;
hatched (dotted) region: the necessary (sufficient) causality conditions
are violated (not satisfied). Bottom: as
Fig.~\ref{Description_experimental_data_1}, for the same four evolutions.}
	\label{MIS_difference_espectra_1}
\end{figure}

We examine in more detail four of the evolutions discussed in the EFT
analysis of the main text, all with the same initial data and parameters:
relativistic Navier--Stokes in frames~\eqref{sharply_causal_frame1}
and~\eqref{sharply_causal_frame2}, and MIS with relaxation times $\tau_\pi$
and $\tau_\pi/2$ (Fig.~\ref{MIS_difference_espectra_1}). Here we compare
their freeze-out surfaces and hadron spectra. The freeze-out surfaces of
the two Navier--Stokes evolutions are indistinguishable, whereas the MIS
surface is wider, and halving $\tau_\pi$ brings it closer to the
Navier--Stokes one (top left). The MIS evolution with $\tau_\pi/2$ (top
right), however, violates the necessary causality conditions over most of
the evolution, showing explicitly that a reduction of second-order effects
comes at the expense of causality. Using the same normalization factors for
the four evolutions, the hadron spectra (bottom) follow the same pattern:
the two Navier--Stokes frames give indistinguishable results, confirming
that the frame dependence is negligible also for observables, while halving
$\tau_\pi$ moves the MIS spectra toward the Navier--Stokes ones, further
confirming that their deviations are second-order effects.


\clearpage

\raggedbottom

\onecolumngrid
\begin{center}
    {\large\bfseries Supplemental Material for:\\
Relativistic Navier--Stokes description of the quark-gluon plasma radial flow}
\end{center}
\twocolumngrid

\noindent
\textbf{Explicit relativistic Navier--Stokes equations.}
For completeness, we give here the explicit form of the relativistic
Navier--Stokes equations, although it is not used explicitly in this
work. Substituting the stress tensor, Eq.~(6) of the main text, into the
conservation equation, Eq.~(2), the relativistic Navier--Stokes
equations in a general frame are
\begin{subequations}
\medmuskip=2mu\thickmuskip=3mu
	\begin{align}
		 \dot{\epsilon}+(\epsilon+p)\nabla \ydot  u  =&-\dot{\mathcal{A}}-((1+c_s^2)\mathcal{A}-\zeta \, \nabla \ydot  u)\nabla \ydot u   \nonumber \\   
		 &-\nabla \ydot  \mathcal{Q}+u \ydot  \dot{\mathcal{Q}} +\frac{1}{2}\eta\, \sigma_{\mu\nu}\sigma^{\mu\nu}\,\,, \label{explicit_EOMsa}\\
		(\epsilon+p)\dot{u}^{\mu}+c_s^2 \nabla_{\perp}^{\mu} \epsilon=&-((1+c_s^2)\mathcal{A}-\zeta \, \nabla \ydot u ) \, \dot{u}^{\mu} \nonumber \\ 
		&-\nabla_{\perp}^{\mu}(c_s^2\mathcal{A}-\zeta\nabla \ydot u)+\Delta_{\lambda}^{\mu}\nabla_{\nu}(\eta\,\sigma^{\nu\lambda}) \nonumber \\
		&-\dot{\mathcal{Q}}^{\mu}- \frac{4}{3}\nabla \ydot u \, {\mathcal{Q}}^{\mu}- u\ydot \dot{\mathcal{Q}}\, u^{\mu} \nonumber \\
		&-\mathcal{Q}_{\nu}(\frac{1}{2}\sigma^{\nu\mu}+\omega^{\nu\mu}) \,\,,\label{explicit_EOMsb} 
	\end{align}
	\label{explicit_EOMs} 
\end{subequations}
where $\omega^{\mu\nu} := \left( \nabla^{\mu}_{\perp} u^{\nu} - \nabla^{\nu}_{\perp} u^{\mu} \right)/2$. \newline

\noindent
\textbf{Detailed description of the numerical code.}
The main ingredients of the numerical code are summarized in the End
Matter of the main text. Each of them is a standard technique for the numerical solution
of hyperbolic partial differential equations with finite differences. For reproducibility and completeness, we describe them here in
detail. 

\textit{Evolution equations.}
We use the coordinates of the main text, $\{\tau,r,\phi,\eta_s\}$. Boost invariance and
azimuthal symmetry imply that all fields depend only on the proper time
$\tau$ and the radial coordinate $r$.  The evolved fields are the energy density $\epsilon(\tau,r)$ and the radial component of the fluid velocity $u^r(\tau,r)$, with $u^\tau=\sqrt{1+(u^r)^2}$ fixed by normalization.
The relativistic Naiver--Stokes equations are second order in derivatives.
Denoting $U=\{\epsilon,u^r\}$, we solve them explicitly for the second time derivatives,
\begin{equation}
  \partial_\tau^2U=F\big(\tau,r,U,\partial_\tau U,\partial_r U,
  \partial_r^2U,\partial_\tau\partial_r U\big)\,,
\end{equation}
Introducing $V\equiv\partial_\tau U$, we obtain a system that is first order in time for the four variables $\{U,V\}$,
\begin{equation}
  \partial_\tau U=V\,,\qquad
  \partial_\tau V=F\big(\tau,r,U,V,\partial_r U,\partial_r^2U,\partial_r V\big)\,.
\end{equation}

The MIS equations are already first order in time, the symmetry
assumptions are the same, and the evolved variables are
$\{\epsilon,u^r,\pi^{rr},r^2\pi^{\phi\phi},\Pi\}$, where we evolve
$r^2\pi^{\phi\phi}$ rather than $\pi^{\phi\phi}$, which
diverges as $r^{-2}$ at the origin.

We explain in detail the techniques and procedure on how to solve these 1+1D PDEs as described in the End Matter. These techniques are standard, and we reproduce them for completeness.

\textit{Grid.}
We discretize the radial direction on a uniform grid that extends from
$r=-R$ to $r=R$, with $R=39.5\,\mathrm{fm}$, passing through the
origin: one point is exactly at $r=0$. An important point in this axially symmetric system is that the equations present a coordinate singularity at $r=0$. This is resolved, as explained in the End Matter, by using l'H\^opital's rule at $r=0$, which is a point that we include in the grid. For these equations, it reduces to expand the equations in Taylor series around $r=0$ and retain the first non trivial term.

The fields are extended to $r<0$ by reflection symmetry, with
$\epsilon$ even and $u^r$ odd, so that periodic boundary conditions can
be imposed at $r=\pm R$. The boundaries lie in the low-density region,
where the fields take their regulator values, and are far enough that no
signal reaches them during the evolution. We use $N=200$, $400$ and
$800$ grid points, with spacing $\Delta r=2R/N$, for the convergence
tests, and $N=800$ for the results shown in the main text. The time step
is $\Delta\tau=C\,\Delta r$, with Courant factor $C=0.1$ for the
Navier--Stokes evolutions and $C=0.02$ for the MIS evolutions.

\textit{Finite differences.}
Spatial derivatives are computed with standard sixth-order centered
finite differences. For a function $f_i=f(r_i)$,
\begin{equation}
  \partial_r^n f\big|_i=\frac{1}{\Delta r^n}\sum_{k=-3}^{3}c^{(n)}_k f_{i+k}
  +\mathcal{O}(\Delta r^6)\,,\qquad n=1,2\,,
\end{equation}
with
\begin{align}
  c^{(1)}_k&=\tfrac{1}{60}\,(-1,\,9,\,-45,\,0,\,45,\,-9,\,1)\,,\\
  c^{(2)}_k&=\tfrac{1}{180}\,(2,\,-27,\,270,\,-490,\,270,\,-27,\,2)\,,
\end{align}
for $k=-3,\dots,3$.
With periodic boundary conditions, the same stencils are used at every
grid point, with indices taken modulo $N$. The mixed derivatives
$\partial_\tau\partial_r f$ are computed by applying the first-derivative
stencil to the evolved time derivatives.
The second-derivative stencil is needed only for the relativistic
Navier--Stokes equations, since the MIS equations involve only first
spatial derivatives of the evolved variables.

\textit{Time integration.}
After the spatial discretization, the equations of motion take the form
$d q /d\tau=\bm{L}(\tau,\bm{q})$, where $\bm{q}$ collects the evolved
variables at all grid points (for the relativistic Navier--Stokes
equations, the fields and their first time derivatives) and $\bm{L}$ is
the discretized right-hand side. We denote $\bm{q}^n=\bm{q}(\tau_n)$ and
$\bm{L}^n=\bm{L}(\tau_n,\bm{q}^n)$, with $\tau_n=\tau_0+n\Delta\tau$.
We integrate in time with the fourth-order Adams--Bashforth (AB4)
method,
\begin{equation}
  \bm{q}^{n+1}=\bm{q}^n+\frac{\Delta\tau}{24}
  \big(55\,\bm{L}^n-59\,\bm{L}^{n-1}+37\,\bm{L}^{n-2}-9\,\bm{L}^{n-3}\big)\,,
\end{equation}
which requires a single evaluation of $\bm{L}$ per time step.

We use a Courant factor $\Delta\tau/\Delta r=0.1$ for the relativistic
Navier--Stokes evolution and $0.02$ for MIS. For MIS, larger values lead to
numerical instabilities, indicating a violation of the
Courant--Friedrichs--Lewy (CFL) condition.

\textit{Initialization of the time integration.}
Since AB4 uses the right-hand side at four consecutive time levels, the
first three steps are computed with the classical fourth-order
Runge--Kutta (RK4) method,
\begin{equation}
  \bm{q}^{n+1}=\bm{q}^n+\frac{\Delta\tau}{6}
  \big(\bm{k}_1+2\bm{k}_2+2\bm{k}_3+\bm{k}_4\big)\,,
\end{equation}
with
\begin{align}
  \bm{k}_1&=\bm{L}\big(\tau_n,\,\bm{q}^n\big)\,,\nonumber\\
  \bm{k}_2&=\bm{L}\big(\tau_n+\tfrac{\Delta\tau}{2},\,\bm{q}^n+\tfrac{\Delta\tau}{2}\bm{k}_1\big)\,,\nonumber\\
  \bm{k}_3&=\bm{L}\big(\tau_n+\tfrac{\Delta\tau}{2},\,\bm{q}^n+\tfrac{\Delta\tau}{2}\bm{k}_2\big)\,,\nonumber\\
  \bm{k}_4&=\bm{L}\big(\tau_n+\Delta\tau,\,\bm{q}^n+\Delta\tau\,\bm{k}_3\big)\,.
\end{align}
From $n=3$ on, the evolution proceeds with AB4. Since both methods are
fourth-order accurate, this initialization preserves the global order
of the scheme.

\textit{Kreiss-Oliger dissipation.}
In finite-difference evolutions of nonlinear partial differential
equations, high-frequency modes that are not resolved by the grid
naturally accumulate, and it is standard to damp them with artificial
dissipation.
We add
Kreiss-Oliger dissipation to the right-hand side of the equations for
the second time derivatives,
\begin{equation}
  \partial_\tau V=F+Q[V]\,,\qquad
  Q[V]_i=-\frac{\sigma_\mathrm{KO}}{\Delta r}\sum_{k=-4}^{4}d_k V_{i+k}\,,
\end{equation}
with
\begin{equation}
  d_k=\tfrac{1}{256}\,(1,\,-8,\,28,\,-56,\,70,\,-56,\,28,\,-8,\,1)
\end{equation}
for $k=-4,\dots,4$, and $\sigma_\mathrm{KO}=0.4$.
The stencil equals
$\Delta r^8\partial_r^8/256$ up to higher-order corrections, so that
$Q=\mathcal{O}(\Delta r^7)$ for smooth solutions and does not affect the
convergence of the scheme. For a Fourier mode $V_i\propto e^{ikr_i}$,
$Q[V]_i=-(\sigma_\mathrm{KO}/\Delta r)\sin^8(k\Delta r/2)\,V_i$, which
damps the modes at the grid scale, $k\Delta r\sim\pi$, while leaving the
well-resolved modes, $k\Delta r\ll1$, essentially unaffected.
For the MIS evolutions, the same dissipation, with the same
$\sigma_\mathrm{KO}$, is applied to each of the five evolved variables.
By reflection symmetry, $u^r$ must vanish at $r=0$. To prevent a slow
drift of $u^r$ away from zero induced by round-off errors, we set the
Kreiss--Oliger term of the $u^r$ equation to zero at $r=0$ only.

\textit{Summary of a time step.}
Advancing the solution from $\tau_n$ to $\tau_{n+1}$ consists of:
\begin{enumerate}
  \item Start from $\{U,V\}$ at $\tau_n$.
  \item Compute their radial derivatives with the finite-difference
        stencils.
  \item Evaluate the right-hand side, including the Kreiss--Oliger
        dissipation.
  \item Advance to $\tau_{n+1}$ with AB4, or with RK4 during the first
        three steps.
\end{enumerate}

\textit{Convergence order.}
Given three evolutions with $N$, $2N$ and $4N$ grid points, we denote
by $q^{(N)}_{A,i}(\tau)$ the evolved variable $A$ at grid point $i$, and
define the convergence order as
\begin{equation}
  \mathcal{Q}(\tau)=\log_2
  \frac{\sum_{A}\sum_{i=0}^{N-1}\big|q^{(N)}_{A,i}-q^{(2N)}_{A,2i}\big|}
       {\sum_{A}\sum_{i=0}^{N-1}\big|q^{(2N)}_{A,2i}-q^{(4N)}_{A,4i}\big|}\,,
\end{equation}
so that the differences are evaluated at the points of the coarsest
grid, which are common to the three grids. For a scheme of order $p$,
the numerical error scales as $\Delta r^p$, so that $\mathcal{Q}=p$. We
use $N=200$, so that the three resolutions are $200$, $400$ and $800$.
Our scheme combines sixth-order finite differences with fourth-order
time integration at fixed Courant factor, so that the global error after
a fixed evolution time is dominated by the time integration and
$\mathcal{Q}=4$ is expected. Since the Kreiss--Oliger dissipation is
$\mathcal{O}(\Delta r^7)$, it does not affect this order. 
The convergence order of our solutions is shown in Fig. 3 of the End Matter.

\textit{Cooper--Frye formula.} For an axially symmetric and boost-invariant
system, the Cooper--Frye formula~\cite{Cooper:1974mv} takes the
form~\cite{Heinz:2004qz}
\begin{align}
&\frac{d^2N}{m_T\,dm_T\,dy} \propto \int r\,dr\,\tau_f \nonumber \\
&\quad\times\Big[m_T K_1\Big(\frac{m_T\sqrt{1+(u^r)^2}}{T_f}\Big)
I_0\Big(\frac{p_T u^r}{T_f}\Big) \nonumber \\
&\qquad - p_T\,\tau_f'\,K_0\Big(\frac{m_T\sqrt{1+(u^r)^2}}{T_f}\Big)
I_1\Big(\frac{p_T u^r}{T_f}\Big)\Big]\,,
\label{Cooper-Frye}
\end{align}
which gives the multiplicity per unit rapidity $y$ and transverse mass
$m_T:=\sqrt{m^2+p_T^2}$, where $p_T$ is the transverse momentum, $K_{0,1}$
and $I_{0,1}$ are modified Bessel functions, and $\tau_f(r)$ and $u^r(r)$
are the proper time and radial velocity on the freeze-out surface. 
The proportionality constant is fixed separately
for each species (pions, kaons and protons), and is the same for the Navier--Stokes and MIS spectra,
since our equation of state assumes chemical equilibrium down to kinetic
freeze-out~\cite{Heinz:2004qz,Heinz:2002un}. We omit the viscous correction
to the distribution function, usually included in viscous
hydrodynamics~\cite{Teaney:2003kp,Jeon:2015dfa}, as well as the hadronic
cascade after freeze-out, which accounts for resonance decays.
Since $\tau_f(r)$ is multivalued, with three branches in the evolutions
considered here, we integrate over the upper branch, from $r=0$ to the
maximum radius of the freeze-out surface. On the remaining part, at earlier
times, particles enter the fluid rather than leave it, which occurs when
$\tau_f'(r)>m_T/p_T$~\cite{Jeon:2015dfa}; we therefore exclude it. Since this
condition does not hold for all $p_T$, this is an approximation, applied
identically to the Navier--Stokes and MIS evolutions.

\textit{Ideal hydrodynamics code.}
For the ideal evolution used in Sec.~5, we use the same infrastructure of the code with
the ideal hydrodynamic equations~(3). These equations are first order in
time, so the initial data consist only of $\epsilon$ and $u^r$, Eqs.~(13).
The equation of state, grid and numerical methods are the same.

\textit{Equation of state and transport coefficients.}
We use the parametrization of the equation of state of the 2014 HotQCD
collaboration~\cite{HotQCD:2014kol},
\begin{equation}
  \frac{p}{T^4}=\frac{1+\tanh\left[c_t(\bar t-t_0)\right]}{2}\,
  \frac{p_{\mathrm{id}}+a_n/\bar t+b_n/\bar t^2+d_n/\bar t^4}
       {1+a_d/\bar t+b_d/\bar t^2+d_d/\bar t^4}\,,
  \label{EoS_QCD}
\end{equation}
with $\bar t=T/(154\,\mathrm{MeV})$, $p_{\mathrm{id}}=95\pi^2/180$,
$\{c_t,t_0\}=\{3.87,0.976\}$, $\{a_n,b_n,d_n\}=\{-8.77,3.92,0.342\}$ and
$\{a_d,b_d,d_d\}=\{-1.26,0.842,-0.047\}$. For the shear and bulk
viscosities we use the JETSCAPE parametrizations~\cite{Parkkila:2021tqq},
\begin{align}
  (\eta/s)(T)&=(\eta/s)_0+(\eta/s)_{\mathrm{s}}\,(T-\hat T_c)\,(T/\hat T_c)^{(\eta/s)_{\mathrm{crv}}}\,,\nonumber\\
  (\zeta/s)(T)&=(\zeta/s)_{\max}\big/\big[1+(T-T_\zeta)^2/w_\zeta^2\big]\,,
  \label{shear_bulk_viscosities_QCD}
\end{align}
with $\hat T_c=0.147$, $(\eta/s)_0=0.104$, $(\eta/s)_{\mathrm{s}}=0.425$,
$(\eta/s)_{\mathrm{crv}}=-0.738$, $(\zeta/s)_{\max}=0.010$,
$T_\zeta=0.170$ and $w_\zeta=0.057$, with $T$ in GeV. 

The parametrizations above are fitted at temperatures well above those
reached in the low-density regulator region, and their extrapolation to
low temperatures is problematic. First, the equation of state reaches
$\epsilon=0$ at $T\simeq54\,\mathrm{MeV}$ with finite pressure, so that
the regulator region, which expands longitudinally, reaches negative
energy densities in a finite time. 
Second, Eq.~\eqref{shear_bulk_viscosities_QCD} gives $\eta<0$
below $T\simeq45\,\mathrm{MeV}$. We therefore modify both below
$T_*=80\,\mathrm{MeV}$, outside the freeze-out surface. For
$\epsilon<\epsilon_*\equiv\epsilon(T_*)$, we use
\begin{equation}
  p=0.3\,\epsilon+B\,\epsilon_*\,(\epsilon/\epsilon_*)^k\,,
\end{equation}
with $B=-0.048$ and $k=1.85$ fixed by continuity of $p$ and $c_s^2$ at
$\epsilon_*$, and obtain the temperature from
$d\ln T=c_s^2\,d\epsilon/(\epsilon+p)$. The speed of sound then
increases smoothly from $c_s^2(T_*)\simeq0.21$ to $0.3$ as
$\epsilon\to0$, and $p\to0$, so that the regulator region never reaches
$\epsilon=0$. For the viscosities, we keep $\eta/(\epsilon+p)$ and $\zeta/(\epsilon+p)$ constant at their values at $T_*$. 


\noindent
\textbf{Physical interpretation of the spectra.}
Although standard in heavy-ion phenomenology, we include for completeness a brief physical interpretation of the spectra in Fig.~2 of the main text, which shows how they are shaped by radial flow.
In the absence of significant radial flow, as in proton--proton
collisions, the Cooper--Frye formula, Eq.~(33), reduces
to $\propto m_\perp K_1(m_\perp/T_\mathrm{f})$. When plotted versus
$m_\perp$, the spectra then have the same shape for all hadrons, a
behavior known as $m_\perp$ scaling~\cite{Heinz:2004qz}. The spectra in
Fig.~2 of the main text are also similar for all species at large
$p_\perp$, where Eq.~(23) gives $\propto e^{-p_\perp/T_\mathrm{eff}}$,
with $T_\mathrm{eff}$ an effective temperature that captures the
blueshift due to the outward expansion of the fluid. At low $p_\perp$,
the spectra of different hadrons differ for two reasons. First, they are
plotted versus $p_\perp$ rather than $m_\perp$. Second, radial flow
flattens the spectra, more strongly at low $p_\perp$ and for heavier
hadrons; this is one of the most distinctive signatures of collective
behavior.

\noindent
\textbf{Effective field theory analysis.}
The ratios of $L_1$ norms given in the main text suffice to assess whether
the evolutions lie within the EFT regime. For the interested reader, here we
show the spatial distributions from which these norms are computed. We use
\begin{equation}
  \|f\|_1(\tau)=\int_{0}^{R}dr\,\big|f(\tau,r)\big|
  \simeq\Delta r\sum_{r_i\geq0}\big|f(\tau,r_i)\big|\,,
\end{equation}
where the sum runs over the grid points.

\begin{figure}[H]
	\centering
	\includegraphics[width=0.4\textwidth]{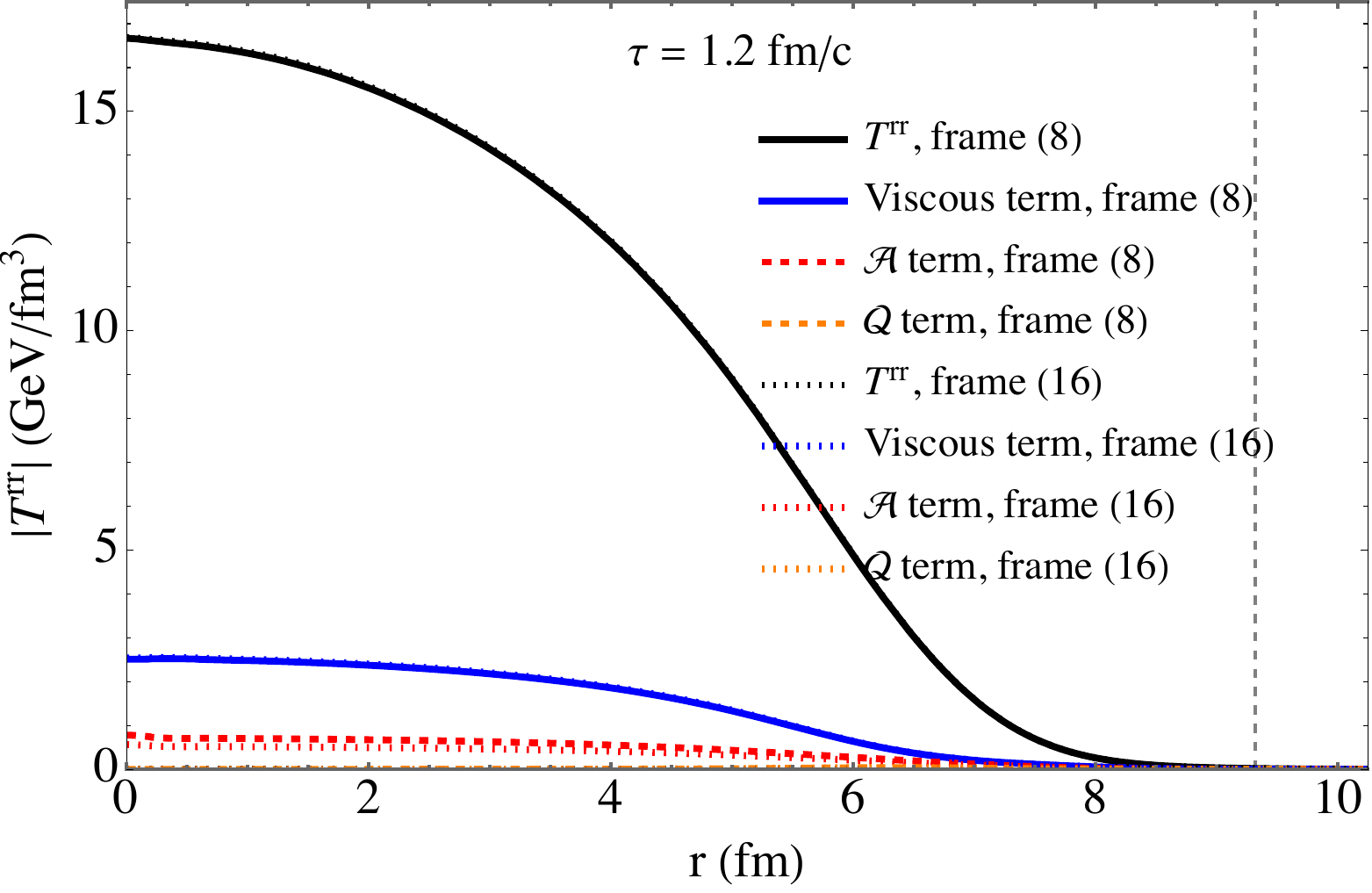}
	\includegraphics[width=0.4\textwidth]{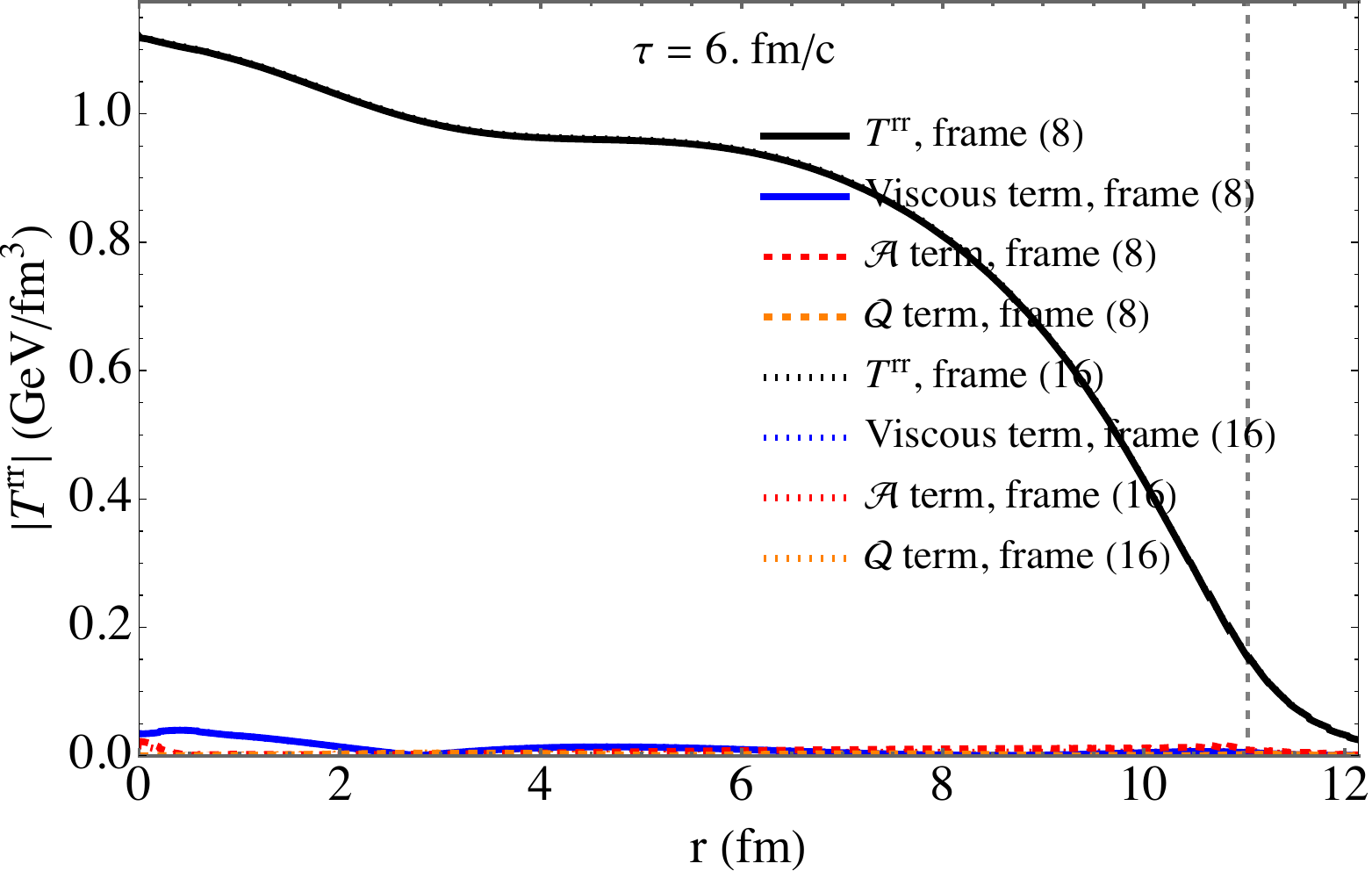}
\caption{Contributions to the stress tensor component $T^{rr}$ as a
	function of $r$ for the evolution in Fig.~1 (left) of the main text, at
	$\tau=1.2$ (top) and $6.0$~fm/$c$ (bottom), in frames~(8) and~(16) of the
	main text: total $T^{rr}$ (black), viscous (blue), $\mathcal{A}$ (red) and
	$\mathcal{Q}^{\mu}$ (orange) terms, for frame~(8) (solid and dashed) and
	frame~(16) (dotted). The reduction of $a_1$ by a factor $16/25$ in
	frame~(16) is visible in the $\mathcal{A}$ term at $\tau=1.2$~fm/$c$,
	whereas the difference in $T^{rr}$ between the two frames is much smaller.
	The vertical dashed line marks the freeze-out radius.}
	\label{NS_EFT_SP}
\end{figure}

Figure~\ref{NS_EFT_SP} shows the contributions to $T^{rr}$ for the
relativistic Navier--Stokes evolution at $\tau=1.2$ and $6.0$~fm/$c$, in
frames~(8) and~(16) of the main text. The viscous term is much smaller than
$T^{rr}$, and decreases in relative size with time. The terms $\mathcal{A}$
and $\mathcal{Q}^\mu$, which regularize the equations, are smaller than the
viscous term, consistent with their being of second order on shell. The
reduction of $a_1$ by a factor $16/25$ in frame~(16) is clearly visible in
the $\mathcal{A}$ term at $\tau=1.2$~fm/$c$, whereas the curves for $T^{rr}$
in the two frames are indistinguishable by eye.

Figure~\ref{MIS_EFT_SP} shows the same quantities for the MIS evolution,
with relaxation times $\tau_\pi$ and $\tau_\pi/2$, together with the
comparison between the evolved dissipative stresses and their first-order
values discussed in the main text. Halving $\tau_\pi$ substantially reduces
the viscous term, which in turn changes $T^{rr}$ appreciably.

\begin{figure}[H]
	\centering
		\includegraphics[width=0.4\textwidth]{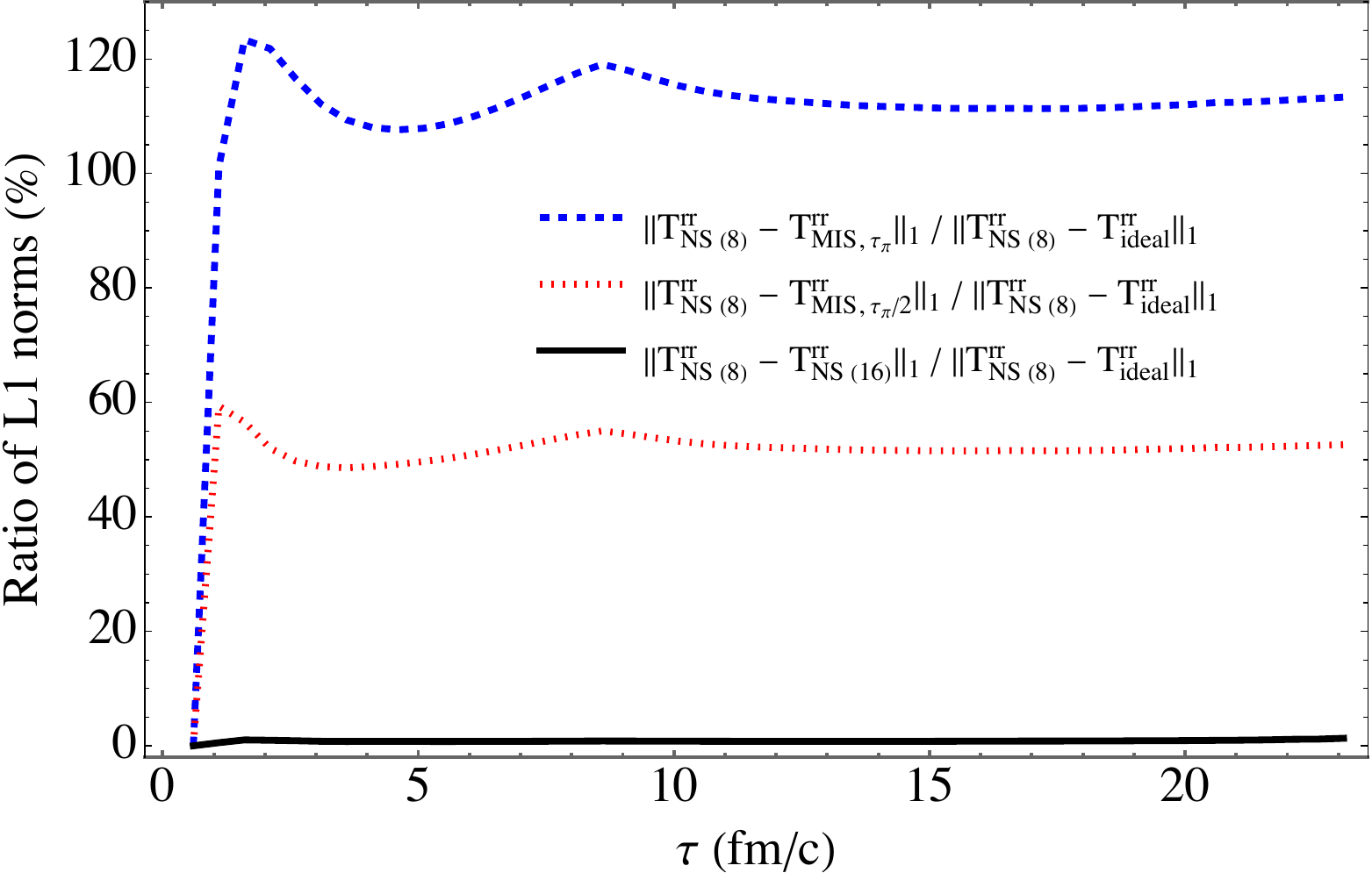}
	\includegraphics[width=0.4\textwidth]{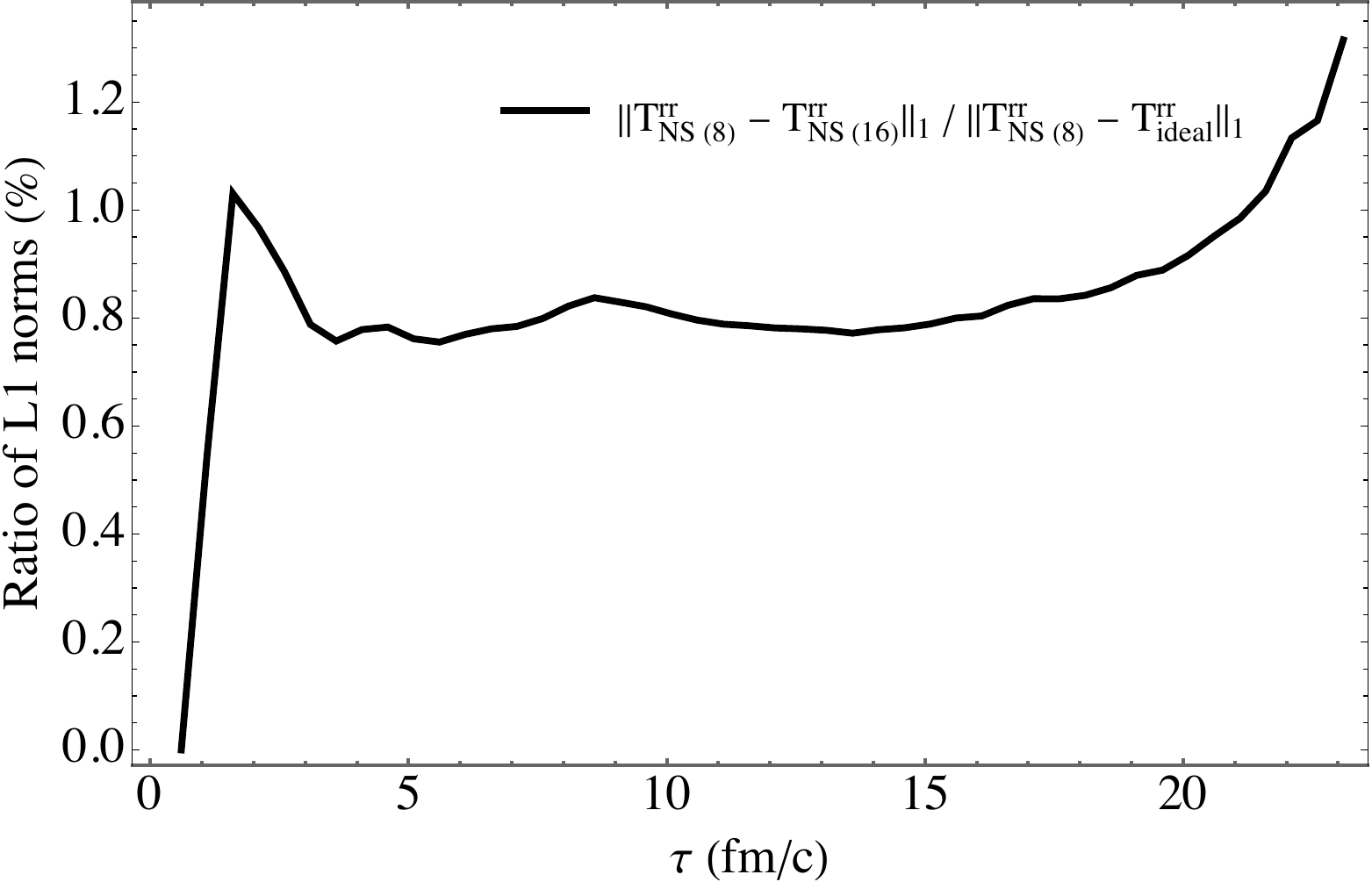}
\caption{Ratios of $L_1$ norms as a function of proper time, normalized to
	the viscous correction of relativistic Navier--Stokes,
	$\|T^{rr}_{\mathrm{NS}(8)}-T^{rr}_{\mathrm{ideal}}\|_1$, where NS(8) and
	NS(16) denote the evolutions of Fig.~1 (left) of the main text in
	frames~(8) and~(16) of the main text. Top: difference between
	relativistic Navier--Stokes and MIS with relaxation times $\tau_\pi$
	(blue dashed) and $\tau_\pi/2$ (red dotted), and frame difference,
	$T^{rr}_{\mathrm{NS}(8)}-T^{rr}_{\mathrm{NS}(16)}$ (black solid). Bottom:
	zoom on the frame difference, which is not visible on the scale of the top
	panel. The maximum values are ${\sim}120\%$ and ${\sim}60\%$
	for MIS with $\tau_\pi$ and $\tau_\pi/2$, respectively, and ${\sim}1\%$
	for the frame difference. All ratios vanish at $\tau_0$, since all
	evolutions start from the same stress tensor and, with Eqs.~(14), the terms
	$\mathcal{A}$ and $\mathcal{Q}^\mu$ vanish there in both frames.}
	\label{fig:criterionC}
\end{figure}

\begin{figure}[H]
	\centering
	\includegraphics[width=0.4\textwidth]{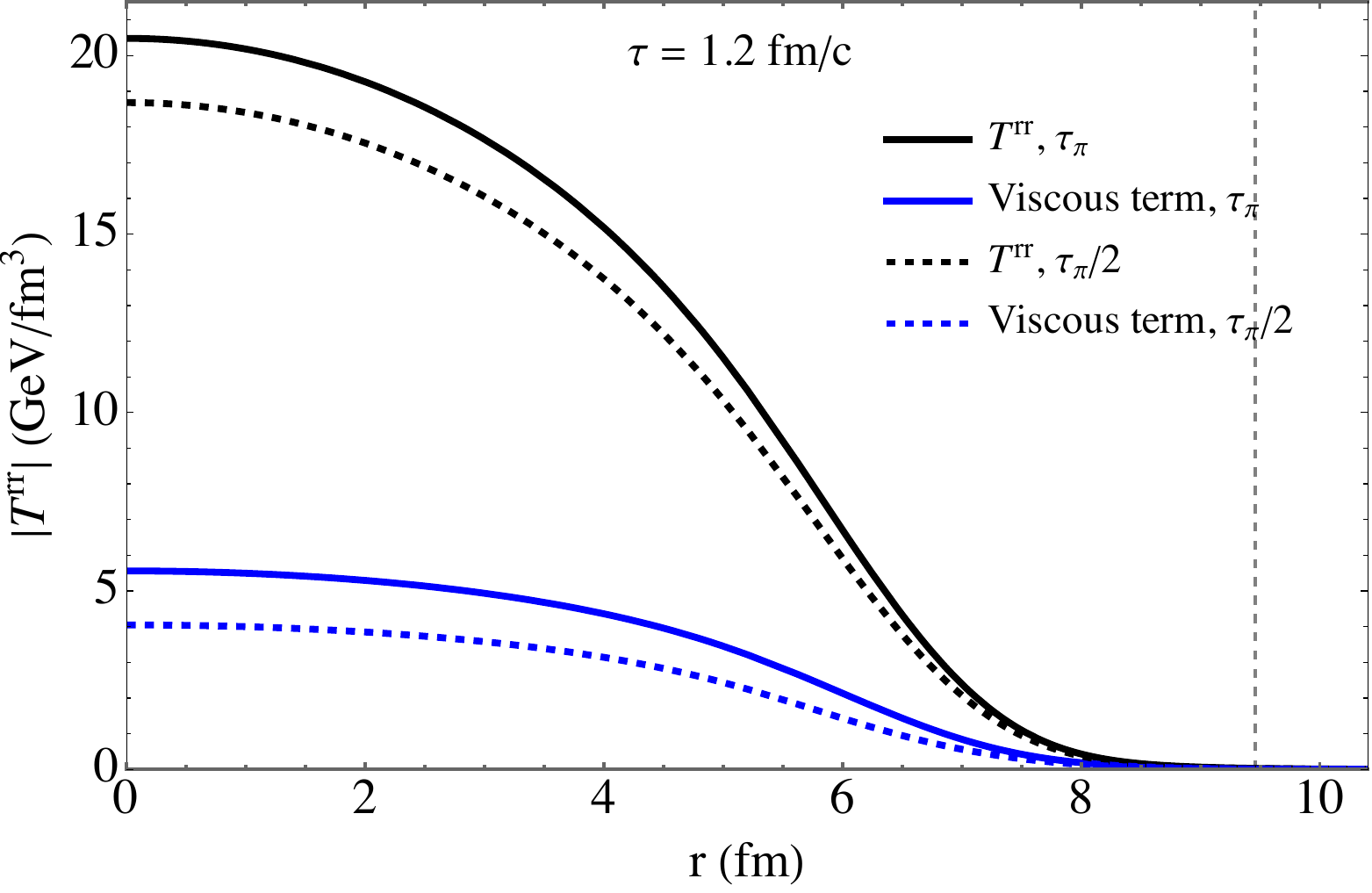}
	\includegraphics[width=0.4\textwidth]{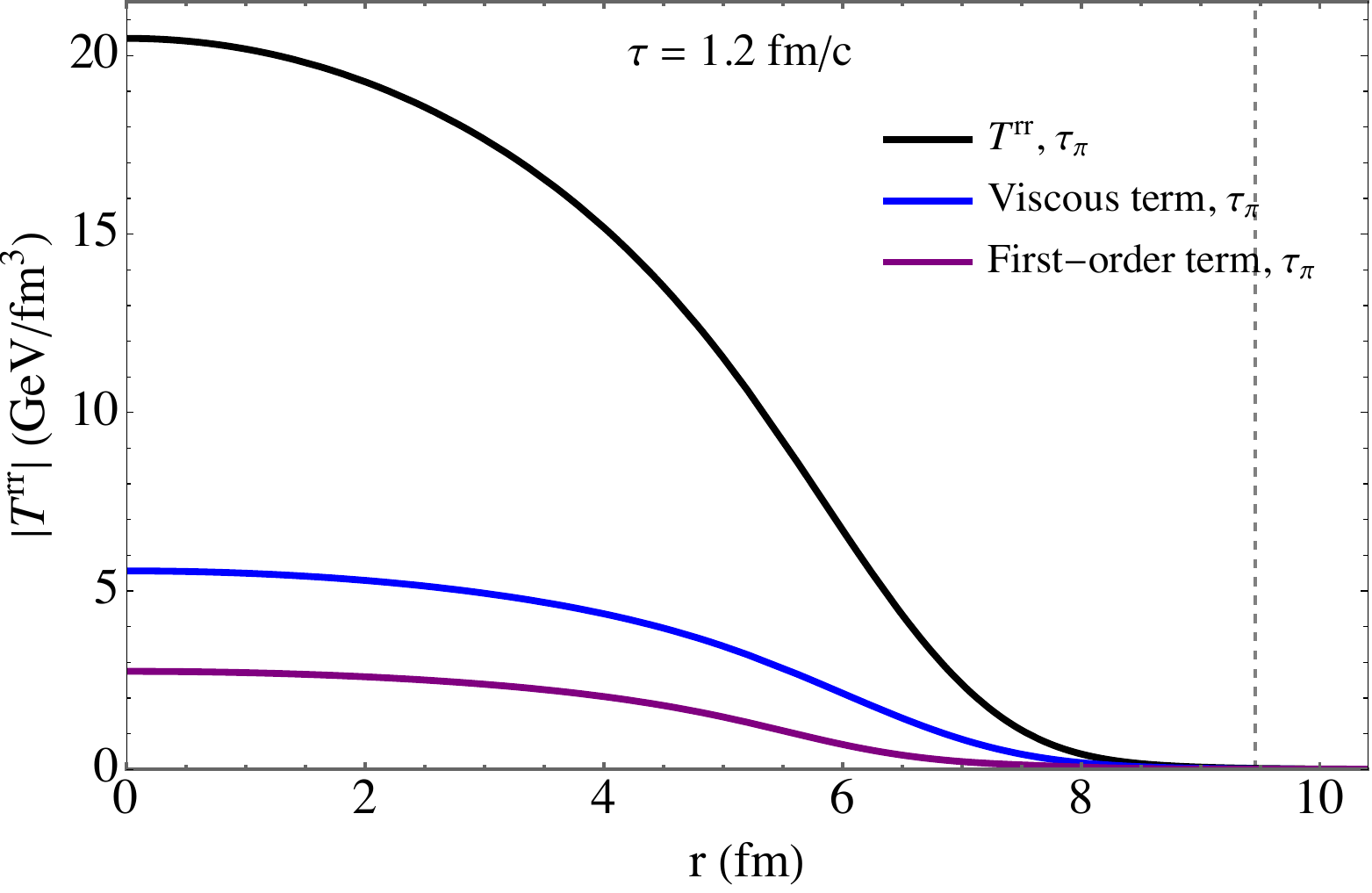}
	\includegraphics[width=0.4\textwidth]{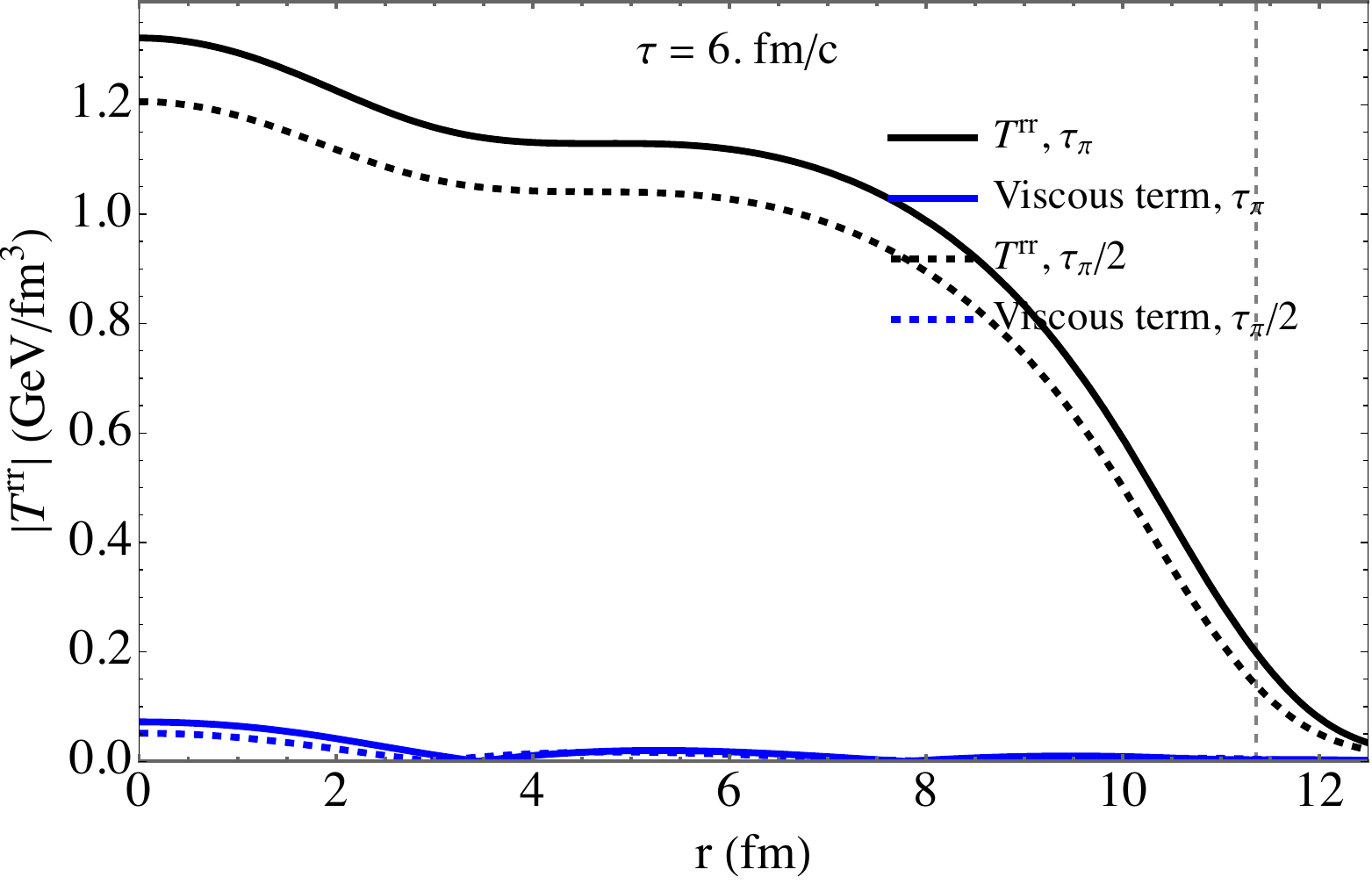}
\caption{Contributions to $T^{rr}$ as a function of $r$ for the MIS
	evolution in Fig.~1 (right) of the main text. Top and bottom: total
	$T^{rr}$ (black) and viscous term, $\pi^{rr}+\Pi\,\Delta^{rr}$ (blue), at
	$\tau=1.2$ and $6.0$~fm/$c$, respectively, for relaxation times $\tau_\pi$
	(solid) and $\tau_\pi/2$ (dashed); the vertical dashed line marks the
	freeze-out radius for $\tau_\pi$. Middle: total $T^{rr}$ (black) and
	evolved viscous term, $\pi^{rr}+\Pi\,\Delta^{rr}$ (blue), compared with
	the first-order value of the latter,
	$-\eta\sigma^{rr}-\zeta\nabla\ydot u\,\Delta^{rr}$, computed from the
	evolved $\epsilon$ and $u^r$ (purple), at $\tau=1.2$~fm/$c$. The
	first-order value is about half the evolved one, showing that
	second-order effects are comparable to first-order ones.}
	\label{MIS_EFT_SP}
\end{figure}

The two comparisons are analogous: in both cases, a parameter that should
only affect higher-order terms is changed by a factor of order one, $a_1$ by
$16/25$ and $\tau_\pi$ by $1/2$. In relativistic Navier--Stokes, $T^{rr}$ is
barely affected, as quantified in Fig.~\ref{fig:criterionC}: the $L_1$ norm
of the frame difference remains at the level of about $1\%$ of that of the
viscous correction at all times. In MIS, instead, $T^{rr}$ changes
appreciably, showing that second-order effects are large and that the
evolution lies outside the EFT regime at early times.

\noindent
\textbf{Early growth of the viscous contribution in MIS.}
The initial growth of the viscous-to-ideal ratio in MIS can be understood in
Bjorken flow of a conformal fluid, $p=\epsilon/3$. Denoting
$\phi\equiv\pi^{xx}=\pi^{yy}$, the equations read
\begin{equation}
\frac{d\epsilon}{d\tau}=-\frac{\epsilon+p-2\phi}{\tau}\,,\qquad
\tau_\pi\frac{d\phi}{d\tau}+\phi=\frac{2\eta}{3\tau}\,,
\end{equation}
with $\tau_\pi=5\eta/(\epsilon+p)$, while Navier--Stokes corresponds to
$\phi=2\eta/(3\tau)$. Taking $T=0.45$~GeV at $\tau_0=0.6$~fm/$c$,
$\eta/s=0.16$ and Navier--Stokes initial data, the ratio $\phi/p$ equals
$31\%$ at $\tau_0$ in both cases. In Navier--Stokes it decreases
monotonically, to $19\%$, $11\%$ and $6\%$ at $\tau=1.2$, $2.8$ and
$6.0$~fm/$c$, whereas in MIS it first grows, to $38\%$ at $\tau=1.2$~fm/$c$,
and then decreases, to $19\%$ and $8\%$. Since initially
$\tau_\pi/\tau\simeq0.6$, $\phi$ cannot follow the rapid decay of its
Navier--Stokes value and remains close to its initial value while the
pressure drops; only once $\tau_\pi/\tau$ becomes small does $\phi$ approach
$2\eta/(3\tau)$. This reproduces the behavior observed in the full
simulations.

\end{document}